\documentclass[10pt]{article}
\usepackage{url}
\usepackage{graphicx}
\usepackage{enumerate}
\usepackage[utf8]{inputenc}
\usepackage{lmodern}
\usepackage[T1]{fontenc}
\usepackage{bm}
\usepackage{diagbox}
\usepackage{amsthm}
\usepackage{textcomp}
\usepackage{float}
\usepackage{hyperref}
\usepackage{multirow}
\usepackage{caption}
\usepackage{amsmath}
\usepackage[normalem]{ulem}
\usepackage{subcaption}
\newcommand{\term}[1]{{\tt #1}\normalsize}
\newcommand{\triple}[1]{{$\langle${\tt #1}$\rangle$}\normalsize}
\usepackage[]{algorithm2e}

\DeclareGraphicsExtensions{.pdf,.png,.jpg}
\begin{document}

\begin{flushleft}
{\Large
\textbf\newline{Spec-QP: Speculative Query Planning for Joins over Knowledge Graphs}
}
\newline
\\
Madhulika Mohanty\textsuperscript{1,*},
Maya Ramanath\textsuperscript{1},
Mohamed Yahya\textsuperscript{2},
Gerhard Weikum\textsuperscript{2}
\\
\bigskip
\bf{1} IIT Delhi, New Delhi, India
\\
\bf{2} Max Planck Institute for Informatics, Saarbr\"{u}cken, Germany
\\
\bigskip
* madhulikam@cse.iitd.ac.in

\end{flushleft}
\date{}

\begin{abstract}
Organisations store huge amounts of data from multiple heterogeneous sources in the form of Knowledge Graphs (KGs). One of the ways to query these KGs is to use SPARQL queries over a database engine. Since SPARQL follows exact match semantics, the queries may return too few or no results. Recent works have proposed \textit{query relaxation} where 
the query engine judiciously replaces a query predicate with similar predicates using
weighted relaxation rules
mined from the KG. The space of possible relaxations is potentially too large to  
fully explore and users are typically interested in only top-$k$ results, so such query engines use top-$k$ algorithms for query processing. However, they may still process all the relaxations, many of whose answers do not contribute towards top-$k$ answers. This leads to computation overheads and delayed response times.

We propose Spec-QP, a query planning framework that speculatively determines which relaxations will have their results in the top-$k$ answers. Only these relaxations are processed using the top-$k$ operators. We, therefore, reduce the computation overheads and achieve faster response times without adversely affecting the quality of results. We tested Spec-QP over two datasets - XKG and Twitter, to demonstrate the efficiency of our planning framework at reducing runtimes with reasonable accuracy for query engines supporting relaxations.

\end{abstract}
 
\section{Introduction}
The availability of immense amounts of digitized data and recent advances in automatic information extraction have made the construction of large Knowledge Bases (KBs) possible. These KBs are typically stored as RDF triples of \triple{spo} where \term{s} is the subject, \term{o} is the object and \term{p} is the predicate. Prominent examples of freely available KBs include YAGO \cite{yago}, DBPedia \cite{dbpedia}, Freebase \cite{freebase}, etc.

These RDF KBs are queried using the SPARQL query language, that, at its core consists of \emph{triple patterns}. For example, the following SPARQL query asks: "Which singers also write lyrics and play guitar and piano?”.

SELECT ?s WHERE\{\\
\texttt{
\hspace*{5mm}?s `rdf:type' <singer>.\\
\hspace*{5mm}?s `rdf:type' <lyricist>.\\
\hspace*{5mm}?s `rdf:type' <guitarist>.\\
\hspace*{5mm}?s `rdf:type' <pianist>}\\
\}

where \term{?s} is a variable to be bound in each of the $4$ triple patterns and to be returned as a result.

 \begin{table}
\centering
\begin{tabular}{ |p{2cm}|p{6cm}|}
\hline
 \textbf{Original} & \textbf{Relaxations}\\ 
 \hline
\term{<singer>} & \term{<vocalist>},\term{<jazz\_singer>}, \term{<artist>}\\
\term{<lyricist>} & \term{<writer>}\\
\term{<guitarist>} & \term{<musician>}, \term{<instrumentalist>}\\
\term{<pianist>} & \term{<percussionist>}\\
 \hline
\end{tabular}
\caption{Example relaxations}\label{tab:relax-eg}
\end{table}

An exhaustive list of such singers in the KB can be computed, but users who issue such queries typically want only the top-$k$, \emph{ranked} results. Ranking of SPARQL query results has been studied before in \cite{naga, elbassuoni@cikm2009, colina} and they typically make use of \emph{scores} for each triple in the KB\footnote{The scores could be based on confidence values, popularity, etc.}. However, a problem that users sometimes face when they issue such queries is \emph{low recall}. That is, the KB may not have $k$ results to return (in some cases, the KB may have \emph{zero} results if one or more of the triple patterns do not have a match). In these cases, it is desirable to \emph{relax} the query by changing one or more of the triple patterns, while ensuring that the query still reflects the original information need. For example, a possible relaxation of the query above is to change the triple pattern \triple{?s `rdf:type' <singer>} to \triple{?s `rdf:type' <vocalist>}. Previous works have dealt with doing these relaxations \emph{automatically} and ranking the corresponding results \cite{query-relax, huang@www2012, poulovassilis@iswc2010, xkg2016}. In this paper, we address the problem of \emph{efficiently evaluating} these relaxed queries.

\paragraph*{Query Processing}

Processing queries and their relaxations to return top-$k$ results is computationally expensive. For example, assuming that every triple pattern in the above query has relaxations as shown in Table \ref{tab:relax-eg}, this would lead to a total of $48$ unique queries (that is, original query, query with one relaxation, query with two relaxations, etc.). A naive method would compute the results to each query, sort the results by score and return the top-$k$. 
\begin{figure}[ht!]
\centering
\includegraphics[height=2.3in]{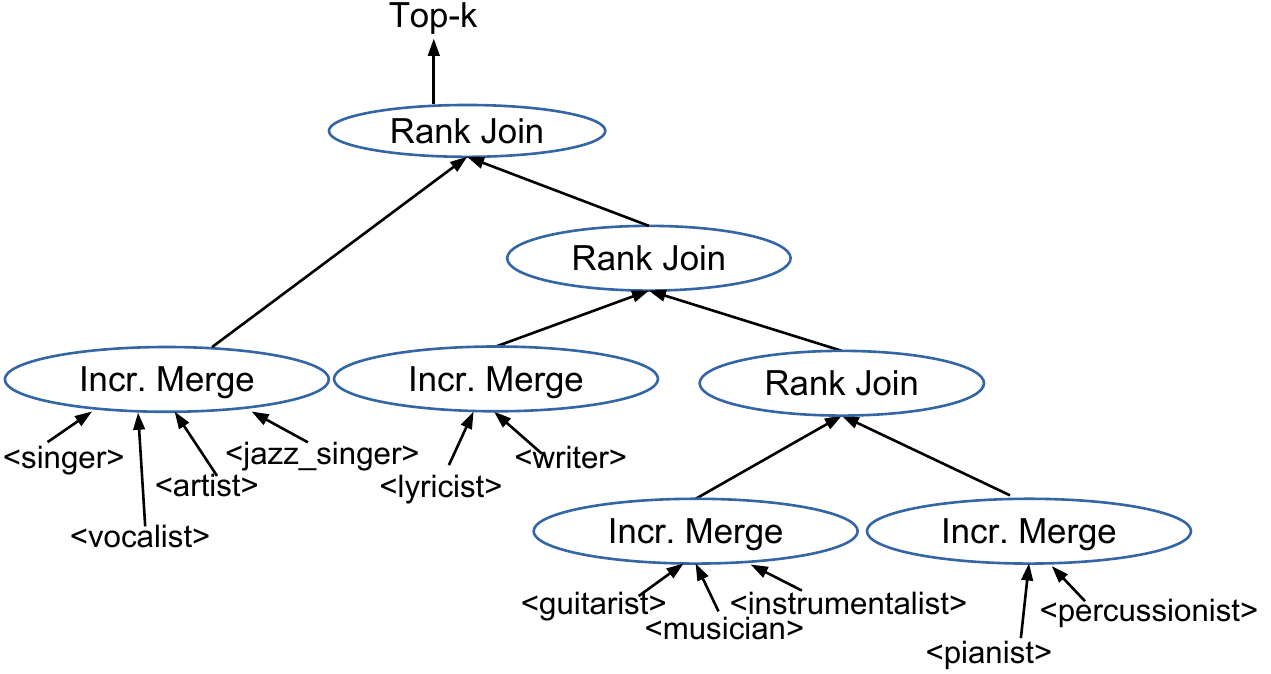}
\caption{Query processing by TriniT. One incremental merge operator is required for each triple pattern and its relaxations. Rank joins are performed over these incremental merges to get top-$k$.}
\label{fig:intro-QP}
\end{figure}

The TriniT \cite{xkg2016} system proposed a mechanism to improve on this naive method. The idea was to compute results from \emph{all} relaxations simultaneously, but in a way that drastically reduced wasteful computations. To this end, TriniT uses two operators: \emph{Incremental Merge} \cite{incmerge2005} (to process the relaxations for a given triple pattern) and \emph{Rank Join} \cite{rankjoin2003} (to compute (partial) join results in sorted order). This is illustrated in Figure \ref{fig:intro-QP}. However, this method still results in wasted resources, since not all relaxations will contribute a result to the top-$k$. For the example query, if we were able to predict that none of the relaxations for the triple patterns with \term{<singer>} and \term{<pianist>} will contribute a result to the top-$k$, then we can replace these incremental merge operators with only ranked joins over the non-relaxed triple patterns.

In this paper, we propose Spec-QP, a speculative approach to prune the space of relaxations resulting in efficient top-$k$ processing of SPARQL queries.  

\paragraph*{Approach and contributions}
Given a KG and a SPARQL query over it, we want to devise an efficient strategy for query processing in the following scenario:
\begin{itemize}
 \item The dataset comprises of triples.
 \item Triples are associated with scores \cite{naga, elbassuoni@cikm2009, colina} and the score of a result is an aggregate of the individual triple scores.
 \item A query can be rewritten using weighted relaxations mined from the KG \cite{query-relax, xkg2016}.
 \item User is interested in only top-$k$ answers.
\end{itemize}

We propose a speculative approach for pruning the space of possible relaxations for a given query. We make use of precomputed statistics about the distribution of scores of the matches to triple patterns in order to speculate on the requirement of relaxations for each triple pattern. This precomputed metadata is an approximation of the score distribution of the answers from the corresponding triple pattern and not the actual scores. When a user enters a query, we estimate the top answer scores that can be achieved using the possible relaxations. This estimation is done using the score distributions and the join cardinality estimates. We then prune those relaxations which are unlikely to contribute triples to the top-$k$ answers based on the top score estimates. Note that our work is orthogonal to any query engine as it can be used on top of any existing graph database engine.

Our main contributions are summarized as follows.
\begin{enumerate}[i.]
 \item A model for the score distribution of individual triple patterns.
 \item A technique to estimate the scores of answers to a query using the above model and using it to predict the presence of answers from each triple pattern's relaxations in the top-$k$.
 \item Pruning the space of relaxations to achieve improved response times over baseline with high prediction accuracy, thereby aiding effective exploration of KGs.
\end{enumerate}

\paragraph*{Organisation}
The rest of the paper is organised as follows: section \ref{sec:definition} introduces some useful definitions and explains the TriniT query processing approach. Section \ref{sec:prob-QP} outlines Spec-QP, the proposed speculative approach to query planning and explains how the plan is executed once the planner generates a plan. Section \ref{sec:expt} summarizes and discusses the experimental results. Section \ref{sec:related} lists the related work and finally section \ref{sec:conclusion} concludes the paper with future work directions.

\section{Preliminaries} \label{sec:definition}
This section introduces some preliminary notions and definitions that will be used henceforth.
  \newtheorem{definition}{Definition}
 
  \begin{definition}\textbf{Knowledge Graphs}\\ 
  Given a set of entities $\mathbf{E}$ and predicates $\mathbf{P}$, a triple $t$ is a tuple $t=$\triple{spo} such that, $t \in \mathbf{E}\times \mathbf{P} \times \mathbf{E}$, $\term{s} \in \mathbf{E}$, $\term{p} \in \mathbf{P}$ and $\term{o} \in \mathbf{E}$. Here, \term{s} is called the ``subject'', \term{p} is the ``predicate'' and \term{o} is the ``object'' of the triple $t$. Each triple is associated with a score, denoted by $S(t)$. These scores represent confidence values or popularity of the triples as previously studied in \cite{naga, elbassuoni@cikm2009, colina}. A set of such tuples can be represented as a graph, which we call a Knowledge Graph, $\mathbf{KG} \subseteq \mathbf{E}\times \mathbf{P} \times \mathbf{E}$.
 \end{definition}
 
 \begin{definition}\textbf{Triple pattern}\\ A triple pattern is of the form $q=$\triple{SPO}, where \term{S}, \term{P} and \term{O} could either be entities or predicates from the KG or variables. Variables are always prefixed with a question mark. A triple pattern matches any triple in the KG having the same values in the designated field. The variables are then bound to the corresponding values in the triple. \end{definition}
  
 \begin{definition}\textbf{Triple pattern query}\\ A triple pattern query is a set of triple patterns, $\mathbf{Q}=\{q_1,q_2,...q_n\}$.
 \end{definition}

 \begin{definition}\textbf{Answer for a Triple pattern query}\\ Given a triple pattern query $\mathbf{Q}$ and a KG, an answer for the query, denoted by $A$, is a mapping of the variables in $\mathbf{Q}$ to values in the KG such that the application of this mapping to each triple pattern $q_i \in \mathbf{Q}$, denoted $A(q_i)$, results in a triple in the KG. The set of all the answers to a query is denoted by the set, $\mathbf{A}$.
 \end{definition}

 \begin{definition}\textbf{Score of a triple matching a triple pattern}\label{def:score-triple}\\ The score of a triple $t$ which matches the triple pattern $q$ is denoted by $S(t|q)$ and is computed as follows:
 
$$S(t|q)=\frac{S(t)}{\max\limits_{t \in \mathbf{A}(q)}(S(t))}$$
 The value ranges between $0$ and $1$.
 \end{definition}

 \begin{definition}\textbf{Score of an answer}\\ The score of an answer $A$ to a query $\mathbf{Q}$ is the aggregation of the scores of the triples resulting from applying the answer mapping to each triple pattern $q_i$ in the query. That is,
 $$ S(A|\mathbf{Q}) = \sum_{q_i \in \mathbf{Q}}S(A(q_i)|q_i)$$
 
 This has been studied previously in \cite{query-relax, huang@www2012, poulovassilis@iswc2010, xkg2016}.

\end{definition}
 \begin{definition}\label{wrr}\textbf{Weighted relaxation rule}\\ A weighted relaxation rule $r$ is a triple $r=(q, q', w)$ where $q$ and $q'$ are triple patterns respectively called the domain and range of the relaxation, and $w\in [0,1]$ denotes the reduction in scores of the triples matching the relaxed triple pattern. Automatic computation of relaxations and the corresponding weights have been studied in \cite{query-relax, xkg2016}.
 
 For example, \triple{?x `rdf:type' <singer>} could be relaxed to \triple{?x `rdf:type' <vocalist>} with a weight of $0.8$, i.e., \\$r=($\triple{?x `rdf:type' <singer>}, \triple{?x `rdf:type' <vocalist>}, $0.8)$.
 \end{definition}

 \begin{definition}\textbf{Relaxed Query}\\ Given a query $\mathbf{Q}$ and a relaxation $r=(q, q', w)$, we say that $r$ applies to $\mathbf{Q}$ if $q \in \mathbf{Q}$. The result of applying $r$ to $\mathbf{Q}$ is a new query $r(\mathbf{Q})=\mathbf{Q'}=(\mathbf{Q} \setminus q)\cup q'$ called the relaxed query.

The score of an answer $A$ obtained through relaxation $r$ applied to a query $\mathbf{Q}$ is defined as: $$S(A|\mathbf{Q'}) = w \times S(A|r(\mathbf{Q})).$$
The score is reduced further for each subsequent relaxation in a similar manner.
Since the same answer could be obtained from multiple relaxed queries, the score of an answer $A$ with respect to the original query and a space of possible relaxations is defined as the maximum score obtained through any relaxation. $$S(A)=\max(S(A|\mathbf{Q'}))$$
\end{definition}

\subsection{Non-Speculative Query Processing (TriniT)}
\label{sec:basicQP}

\begin{figure}[ht!]
\centering
\includegraphics[height=1.5in]{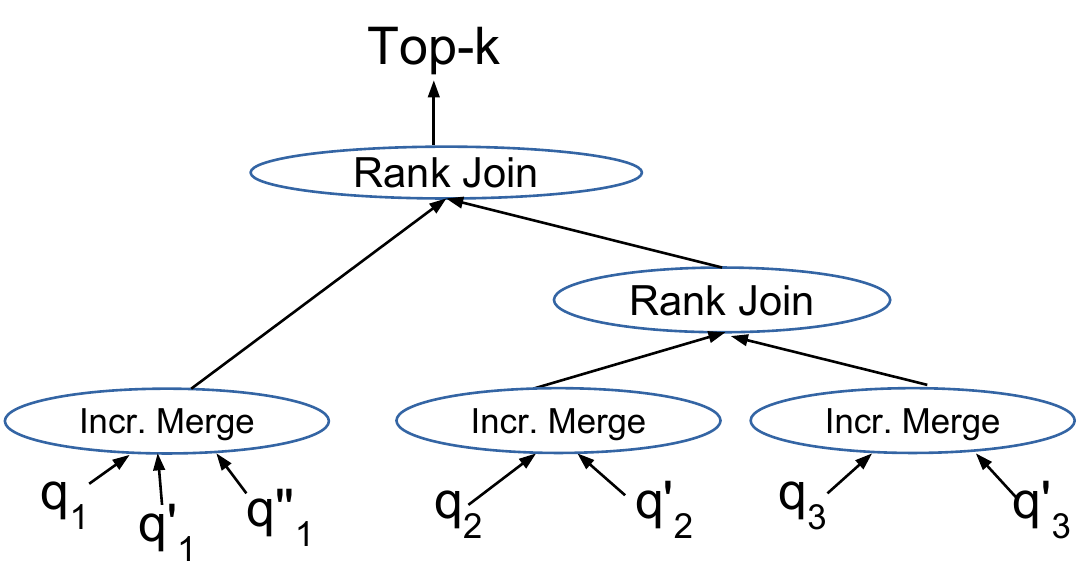}
\caption{Query plan generated by TriniT for the query $\mathbf{Q}=\{q_1,q_2,q_3\}$. One incremental merge operator is required for each triple pattern and its relaxations. A rank join operator takes in two sorted lists and produces a ranked list of (partial) answers from the join.}
\label{fig:naive-QP}
\end{figure}

As mentioned in the Introduction, TriniT computes results from \emph{all} relaxations simultaneously using two operators: \emph{Incremental Merge} \cite{incmerge2005} and \emph{Rank Join} \cite{rankjoin2003}.  
Given the query $\mathbf{Q}=\{q_1,q_2,q_3\}$, and the relaxations, $r_1=(q_1, q'_1, w_1)$,  $r_2=(q_1, q''_1, w_2)$, $r_3=(q_2, q'_2, w_3)$ and $r_4=(q_3, q'_3, w_4)$, Figure \ref{fig:naive-QP} shows the query plan generated by TriniT.
Incremental Merge is used to efficiently scan the list of matches to a triple pattern and all its relaxations to output only one merged and sorted list for each triple pattern. Each of the three incremental merge operators in the example takes as inputs the sorted lists of matches\footnote{Recall that each triple is associated with a score.} for each triple pattern, $q_1$, $q_2$ and $q_3$ and their relaxations. Each of them outputs a combined sorted list of triples for each triple pattern along with its relaxations. The rank join computes a join of the two sorted inputs in an incremental manner until enough results have been produced, while minimising the number of answers read from each list to get top-$k$ answers. This helps avoid computing the entire join and then sorting over it. The inputs for Rank Joins are either the outputs of Incremental Merges or Rank Joins. Both operators use priority queues for already seen answers and maintain upper bounds to estimate scores of the answers that can be obtained by reading further into the lists at any given point. This avoids accessing entire lists of (partial) answers and aids early termination. 

However, TriniT still processes relaxations from all the triple patterns, many of which do not contribute triples towards the top-$k$ answers. Our technique aims to eliminate this inefficiency.

\section{Spec-QP, the Speculative Framework for optimizing Query Plans}\label{sec:prob-QP}
The Incremental Merge and Rank Join operators were introduced by TriniT to flexibly perform relaxations without the need to fully explore the space of all possible answers.  
Note that in the absence of relaxations, we could have simply resorted to rank joins over sorted answer-lists of only the original triple patterns. These joins are straight-forward and much faster than processing the triple pattern and its relaxations using incremental merges and joining over them.

We propose Spec-QP, a query planning approach which uses a predictor to predict whether the relaxations of a triple pattern are likely to be required for producing the top-$k$ answers. We need not process relaxations for those triple patterns whose relaxations are predicted to be not required. The predictor uses an expected score estimator based on the precomputed statistics about the distribution of the scores for triple pattern matches. We first describe the  estimator and then give details of the planning approach.

\subsection{Expected score estimator}\label{sec:exp-pred}
\begin{figure}[ht!]
\centering
\includegraphics[scale=0.55]{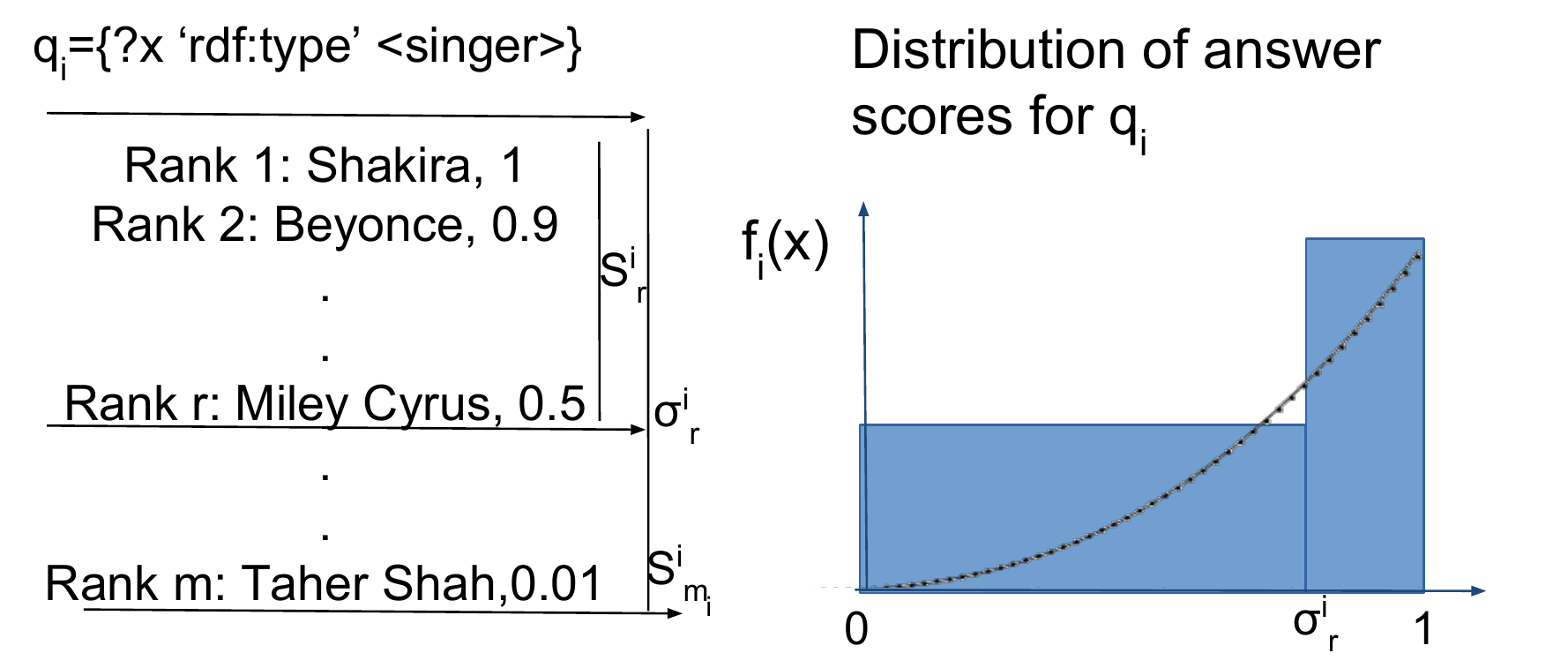}
\caption{Score distribution for answers of a triple pattern modelled as a two-bucket histogram.}

\label{fig:score-dist}
\end{figure}
The expected score estimator is based on order statistics and estimates the expected scores at given ranks for the original as well as relaxed queries. These are used by the query planner to predict the presence of answers from a relaxation in top-$k$.\\
The $m$ matching triples for a triple pattern have scores represented by 
the independent and identically distributed random variables $X_{i1}, X_{i2}, ..., X_{im}$, each with a common distribution, $f_i(x)$. Here, $f_i(x)$ is the probability distribution for the scores of the answers for a triple pattern (or relaxation), $q_i$, from the $KG$. The cumulative distribution function (cdf) is represented by $F_i(x)$. The set \{$X_{i1}, X_{i2}, ..., X_{im}$\} is a sample of size $m$ taken from the distribution $F_i(x)$. The set of the observed values of answer scores \{$x_{i1}, x_{i2}, ..., x_{im}$\} of random variables $X_{i1}, X_{i2}, ..., X_{im}$ is called a
realization of the sample. 
$X_{i(1)}, X_{i(2)}, ..., X_{i(m)}$ are random variables resulting from arranging 
the values of each of $X_{i1}, X_{i2}, ..., X_{im}$ in increasing order, and $X_{i(j)}$ is
called the $j^{th}$ order statistic. Given these random variables and their distributions, we need to estimate the score distribution for the answers of the query, $\mathbf{Q}$. $X_{Q1}, X_{Q2}, ..., X_{Qn}$ are the random variables representing the scores of the $n$ answers to the query, $\mathbf{Q}$ (possibly composed of a single 
triple pattern). $X_{Q(1)}$ is the first order statistic 
corresponding to the lowest scoring answer among all the $n$ answers of $\mathbf{Q}$, and 
$X_{Q(n)}$ is the $n$-th (or largest) order statistic corresponding to the highest scoring answer (ranked 1). A relaxed answer would appear in top-$k$ only when its expected highest score ($X_{Q'(n')}$) amongst its $n'$ answers exceeds the expected $k^{th}$ highest score of the original query ($X_{Q(n-k)}$). In order to compute the expected value at a given rank, we use the result given in \cite{david2004order}: For i.i.d. random variables, $X_{1}, X_{2}, ..., X_{m}$ each with a common distribution, $f(x)$, the expected value of $i^{th}$ order statictic, $X_{(i)}$ can be approximated as $E(X_{(i)}) \approx F^{-1}(\frac{i}{m+1})$ where $F(x)$ denotes the cdf and $m$ is the size of the sample. Using this, the expectation of $X_{Q(i)}$ can be approximated as $E(X_{Q(i)}) \approx F_Q^{-1}(\frac{i}{n+1})$ where $F_Q(x)$ denotes the cdf of the scores for the answers to the query and $n$ is the no. of answers of $\mathbf{Q}$ .

We now give the details of the construction of the probability density function (pdf) of these random variables. 

\subsubsection{Score Distributions for the Triple Patterns:}

For every triple pattern $q_i$ in the $KG$, we store the following precomputed statistics about the scores $\boldsymbol{\sigma^{i}}$ of the matching triples:
\begin{itemize}
 \item $m_i$: the total number of triples matching the triple pattern.
 \item $S_{r}^i$: the cumulative scores of the answers over all the ranks $1$ through $r$ for $r=r_1,r_2,...r_{n-1},m_i$. As described later, these values of $r$ will represent the ranks which form the bucket boundaries for the histograms of the score distributions.
 \item $\sigma_{r}^i$: the scores at rank $r$ for $r=r_{1},r_{2},...r_{n-1}$.
\end{itemize}

We now estimate the scores distribution for answers to triple pattern $q_i$. Note that the ranks will not be explicitly reflected here, it is just the distribution of the answer score values from which each score in \{$X_{i1}, X_{i2}, ..., X_{im}$\} is assumed to be independently sampled. $f_{i}(x)$ and $F_{i}(x)$ are used to denote the pdf and cdf respectively.

The pdf can be modelled as a $n$-bucket histogram in the following way:
\begin{align}
 f_{i}(x) =& \frac{S_{m_i}^i-S_{r_{n-1}}^i}{S_{m_i}^i}\frac{1}{\sigma_{r_{n-1}}^i} \text{ for } 0\leq x< \sigma_{r_{n-1}}^i \\
 & \frac{S_{r_{p}}^i-S_{r_{q}}^i}{S_{m_i}^i}\frac{1}{\sigma_{r_q}^i-\sigma_{r_p}^i} \text{ for } \sigma_{r_{p}}^i\leq x< \sigma_{r_{q}}^i \text{ where } p=q+1\\
 & \frac{S_{r_1}^i}{S_{m_i}^i}\frac{1}{1-\sigma_{r_1}^i} \text{ for } \sigma_{r_1}^i\leq x\leq 1
\end{align}

The pdf is essentially uniform distribution in each bucket with the height being proportional to the score mass in the bucket. 

In order to find the best fitting number of buckets, we observed the scores of the answers to few random triple patterns sorted in decreasing order. We found that they had a power law distribution. The power law distribution follows the $80/20$ rule which states that $80$\% of the score mass lies in the $20$\% of the answers. We, therefore, chose two-bucket histograms to represent the score distributions (as shown in Figure \ref{fig:score-dist}). The short and tall bucket represents the interval which has $80$\% of the score mass. The longer bucket represents the long tail having only $20$\% of the score mass.  We store only the following $4$ values for each triple pattern:
\begin{itemize}
 \item $m_i$: the total number of triples matching the triple pattern.
 \item $\sigma_{r}^i$: the score of the answer at rank $r$ where $r$ represents the rank within which $80$\% of the score mass is contained for the triple pattern matches.
 \item $S_{r}^i$: the cumulative score of the answers over all the ranks $1$ through $r$.
 \item $S_{m_i}^i$: the cumulative score of the answers over all the ranks $1$ through $m_i$.
\end{itemize}

The pdf for each distribution is following:
\begin{align*}
 f_{i}(x) =& \frac{S_{m_i}^i-S_{r}^i}{S_{m_i}^i}\frac{1}{\sigma_{r}^i} \text{ for } 0\leq x< \sigma_{r}^i \\
 & \frac{S_{r}^i}{S_{m_i}^i}\frac{1}{1-\sigma_{r}^i} \text{ for } \sigma_{r}^i\leq x\leq 1
\end{align*}
This pdf gives us the following cdf:

\begin{align*}
 F_{i}(x) =& ax \text{ for } 0\leq x< \sigma_{r}^i \\
 & bx+c \text{ for } \sigma_{r}^i\leq x\leq 1 \\
 \text{with}\\
 a = &\frac{S_{m}^i-S_{r}^i}{S_{m}^i}\frac{1}{\sigma_{r}^i} \text{ and }  \\
 b = &\frac{S_{r}^i}{S_{m}^i}\frac{1}{1-\sigma_{r}^i} \text{ and } \\
 c = &\frac{S_{m}^i-S_{r}^i}{S_{m}^i} - \frac{S_{r}^i}{S_{m}^i}\frac{\sigma_{r}^i}{1-\sigma_{r}^i}
\end{align*}

\subsubsection{Score Distribution for the Triple Pattern Query:}
\begin{figure}[ht]
\centering
\includegraphics[scale=0.55]{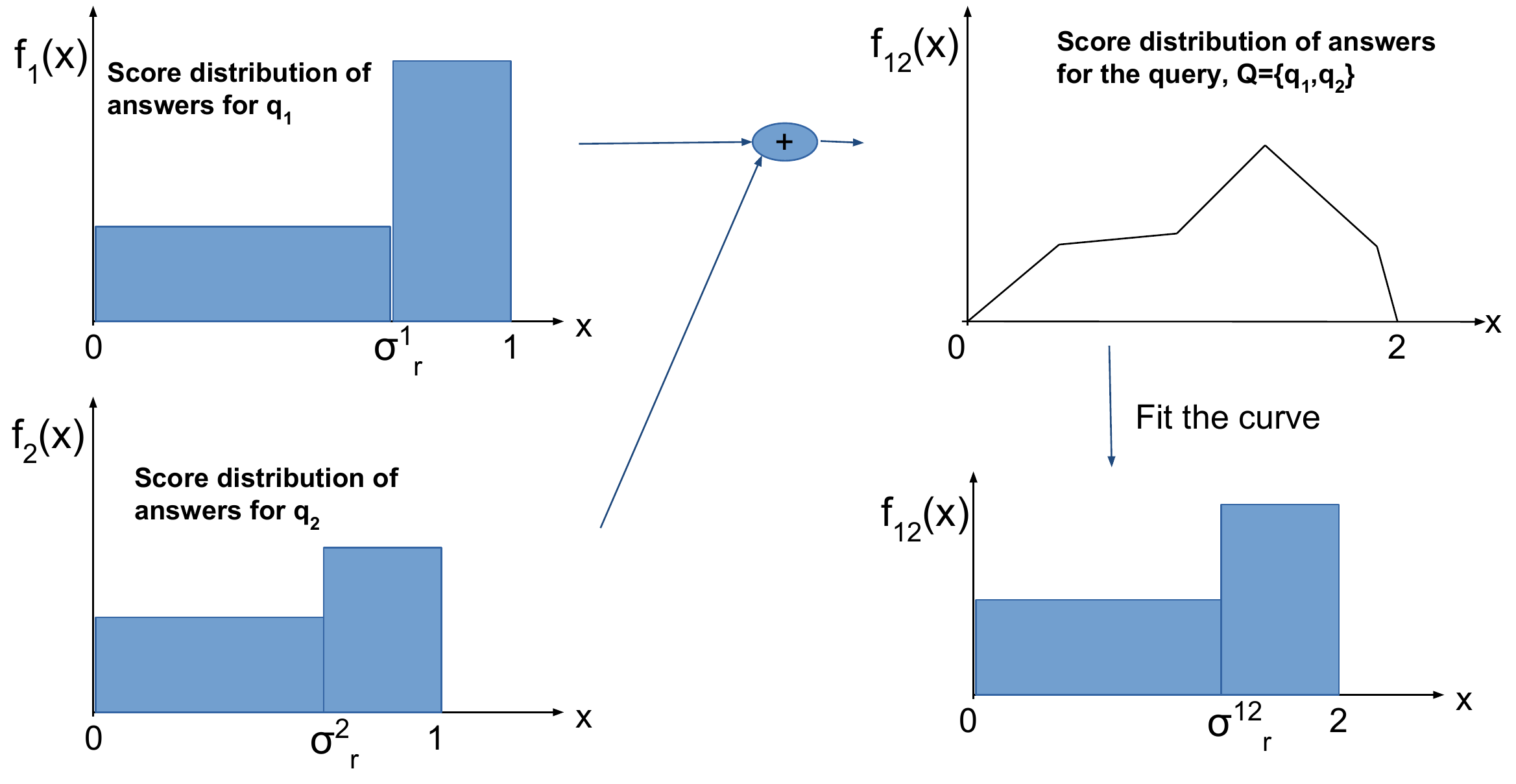}
\caption{Score distribution for a triple pattern query is computed as the convolution of the pdf's of the constituent triple patterns.}
\label{fig:conv}
\end{figure}
The score of an answer for the triple pattern query is the sum of the scores of the individual triples in the answer. Since each triple is contributed by one triple pattern in the query and we have estimates for their scores, we can estimate the scores for answers to the query using the following approach.

Let us assume our triple pattern query, $\mathbf{Q}=$\{$q_1,q_2$\}.  \{$X_{11}$, $X_{12}$, ..., $X_{1m}$\} represents the $m$ triples matching $q_1$ and \{$X_{21}$, $X_{22}$, ..., $X_{2m'}$\} represents the $m'$ triples matching $q_2$. The scores for triples matching these triple patterns have the distributions $f_{1}(x)$ and $f_{2}(x)$ respectively, as defined before. The scores of $\mathbf{Q}$'s $n$ answers are represented by the random variables $X_{Q1}, X_{Q2}, ..., X_{Qn}$. Each of these is a sum of two random variables, one from \{$X_{11}, X_{12}, ..., X_{1m}$\} and another from \{$X_{21}, X_{22}, ..., X_{2m'}$\}. The pdf for the sum of the random variables is given by the convolution of the two individual pdfs, $f_{1}*f_{2}(x)$ (as depicted in Figure \ref{fig:conv}). Hence, the pdf for the scores of the answers to the query is given by the convolution of the pdf's of the scores for matches to the constituent triple patterns. The resulting pdf is a multi-piece-wise linear function. Given the number of results in the combined distribution, $m_{12}$, we can estimate $\sigma_{r}^{ij}$, $S_{r}^{ij}$ and $S_{n}^{ij}$ using the expected score computation from order statistics. This again results in a two-bucket histogram for the distribution of the scores of the answers to the query. For the computation of $m_{12}$, we use the estimates for join selectivity\footnote{Traditional database systems use multiple heuristics to estimate join selectivity. For the purpose of this work, we have taken exact join selectivity values.}, $\phi_{12}$ as $m_{12} = m*m'*\phi_{12}$. For three of more triple patterns, we repeat the above process the required number of times to get the final histogram representing the score distribution for the query.

\subsubsection{Score prediction}\label{subsec:scorepred}
Once we have constructed the pdf and cdf representing the scores for the answers of a given query, we can estimate the expected score, $X_{Q(n-i)}$\footnote{Note that it is $n-i$ and not $i$ since the $n^{th}$ order statistic represents the highest value with rank $1$.}  at a given rank $i$ as $E(X_{Q(n-i)}) \approx F_Q^{-1}(\frac{n-i}{n+1})$ where $F_Q(x)$ denotes the cdf of the query answer scores and $n$ is the no. of answers for $\mathbf{Q}$. Given these estimates for scores at various ranks, we now generate the query plan.

\subsection{Query Planning}\label{sec:QP}
\textbf{Query Plan:} Given a query $\mathbf{Q}$, a query plan consists of subsets of triple patterns $\mathbf{Q_1},\mathbf{Q_2},....,\mathbf{Q_s}$ where
\begin{enumerate}[i.]
 \item each $\mathbf{Q_i}$ consists of one or more triple patterns from $\mathbf{Q}$,
 \item the $\mathbf{Q_i}$'s are pairwise disjoint, and
 \item the union of $\mathbf{Q_i}$'s equals $\mathbf{Q}$.
\end{enumerate}
For example, a query plan for the query $\mathbf{Q}=\{q_1,q_2,q_3\}$, will be $\{\{q_1,q_3\},\{q_2\}\}$. The singletons correspond to the triple patterns which require relaxations.

\begin{algorithm}[t]
\DontPrintSemicolon 
\KwIn{The query $\mathbf{Q}=\{q_1,q_2,...q_n\}$.}
\KwOut{The query plan, $\mathbf{QP}$}
$\mathbf{QP} \gets \{\{\mathbf{Q_1}\}\}$, where $\mathbf{Q_1}=\mathbf{Q}$\;
$f_{\mathbf{Q}}(x) \gets f_1*f_2*..*f_n(x)$\;
Get $E_\mathbf{Q}(k)$ from ``expected score estimator''.\;
\For{$q_i \in \mathbf{Q}$} {
  $q'_i \gets$ top-weighted relaxation for $q_i$\;
  $\mathbf{Q}'\gets \mathbf{Q}-\{q_i\}\cup\{q_i'\}$\;
  $f_{\mathbf{Q'}}(x) \gets f_1*f_2*...*f'_i*..*f_n(x)$\;
  Get  $E_{\mathbf{Q'}}(1)$ from ``expected score estimator''.\;
  \If{$E_{\mathbf{Q'}}(1) > E_\mathbf{Q}(k)$} {
   $\mathbf{QP}=\{\{\mathbf{Q_1}\}-\{q_i\},\{q_i\}\}$\;
  }
}
\Return{$\mathbf{QP}$}\;
\caption{{\sc PLANGEN} generates the query plan.}
\label{algo:max}
\end{algorithm}
\subsubsection{Query plan generation}
The key idea behind the planning approach is that the answers from all of the triple patterns' relaxations do not appear in the top-$k$ answers. We save on computations over such triple patterns by never processing their relaxations. For each triple pattern, only the top-weighted relaxation has the highest top score due to normalization of scores as per Definition \ref{def:score-triple}, i.e, the top score from each relaxation is equal to its weight. Hence, we need to check only the top-weighted relaxation for each triple pattern for its potential to contribute answers towards top-$k$.

Given a query $\mathbf{Q}$ and the score distribution for each triple pattern, the query plan is generated as outlined in Algorithm \ref{algo:max}. {\sc PLANGEN} first predicts the requirement of relaxations for each triple pattern. For prediction, the query planner uses an ``expected score estimator'' described in Section \ref{sec:exp-pred}, which gives estimates of the expected scores at $k^{th}$ rank for the original query, $E_\mathbf{Q}(k)$ and top rank for the highest weighted relaxed query, $E_\mathbf{Q'}(1)$ (for a given triple pattern at a time). If the topmost score from the relaxed query obtained by relaxing a given triple pattern exceeds the $k^{th}$ score from the original query, it predicts that the triple pattern's relaxations are required. Note that our estimator takes into account join score distributions and join cardinalities for estimating the expected score for a given query.

The query plan, $\mathbf{QP}$ returned by {\sc PLANGEN} will have only one subquery, $\mathbf{Q_1}$ of size > $1$, called the ``join group'' (non-relaxed triple patterns), the rest will be only singletons (triple patterns to be relaxed). 

\subsubsection{Query Execution}\label{sec:QE}

Given a speculative query plan $\mathbf{QP}=\{\mathbf{Q_1},\mathbf{Q_2},..,\mathbf{Q_s}\}$ with $s$ subsets generated by the speculative query planner, we execute it in the following manner. 
\begin{enumerate}
 \item The join group, $\mathbf{Q_1}$ is executed as (left-deep) rank joins over the answer lists (sorted by score) for each triple pattern. Note that, none of the triple patterns in this group are relaxed.
 \item The singletons are processed by Incremental Merge operator for each.
 \item Rank joins are performed over the join group and singletons.
\end{enumerate}

\begin{figure}[t!]
\centering
\includegraphics[height=1.2in]{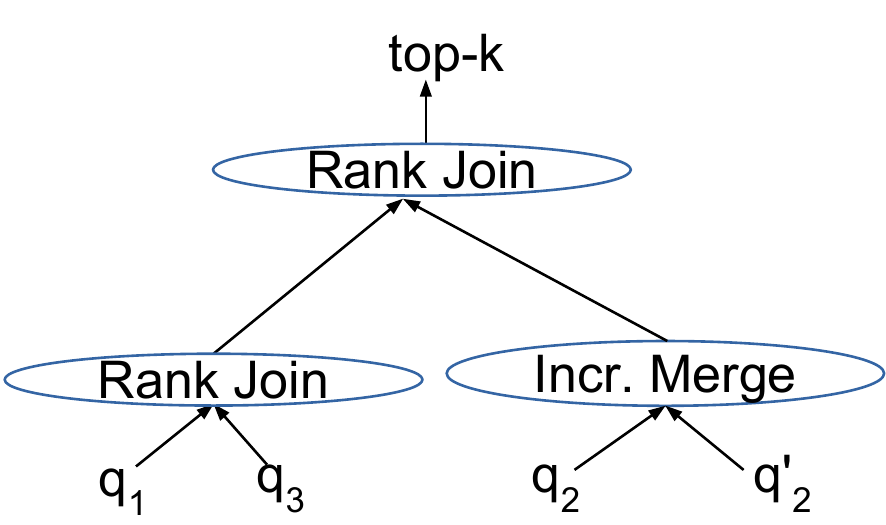}
\caption{Query Plan when $\mathbf{Q}=\{q_1,q_2,q_3\}$ and only $q_2$'s relaxations are predicted to be in top-$k$. Only $q_2$ requires an incremental merge. $q_1$ and $q_3$ are joined using a rank join over the sorted answer lists for each of them. One rank join is required to join these results.}
\label{fig:prob-QP}
\end{figure}

Given $\mathbf{Q}=\{q_1,q_2,q_3\}$, when we predict that $q_1$ and $q_3$ are not going to be relaxed, the effective query plan is $\{\{q_1,q_3\},\{q_2\}\}$.  We use rank joins to compute the join between sorted lists of matches for $q_1$ and $q_3$ and require incremental merge only for $q_2$ and its relaxations. The results from these are joined using Rank Joins to get the final top-$k$ answers. Note that we reduce the number of incremental merges required to $1$ as compared to $3$ by TriniT. This would lead to less computation at run time and thus, faster response times will be achieved. Figure \ref{fig:prob-QP} illustrates this approach. 

The equivalent TriniT plan for this query will be $\{\{q_1\}$, $\{q_2\}$, $\{q_3\}\}$, i.e., all triple patterns occur as different subsets and each of them are processed by Incremental Merges followed by Rank Joins over all these incremental merges (Refer Figure \ref{fig:naive-QP}).

\newcommand\abs[1]{\left|#1\right|}
\section{Experimental Evaluation}\label{sec:expt}
This section discusses the experimental evaluation performed for demonstrating the performance of the speculative planner.

\subsection{Baseline}
We test our system for faster response times with good accuracy in a querying platform which has to give top-$k$ results in the scenario where the user query is allowed to be relaxed to get results that satisfy the desired information need. We compare Spec-QP with the non speculative query processing engine (TriniT) (refer Section \ref{sec:basicQP}) which involves Incremental Merges for relaxations and Rank Joins for joins. Note that TriniT processes \textit{all} the relaxations and outputs the true top-$k$.

\subsection{Datasets used}
We have used two datasets for the purpose of demonstrating the performance of the speculative planner. They are as follows:
\begin{enumerate}
 \item Extended Knowledge Graphs (XKG):\\
  We have used the eXtended Knowledge Graphs (XKG) introduced by TriniT \cite{xkg2016}. This is a RDF format dataset but unlike standard RDF format, the triples are composed of a mixture of textual tokens and IRIs. These ``textual'' content triples are constructed from a document corpus by using OpenIE techniques and NED. Each triple score is equal to the number of times this particular triple was encountered. This knowledge base along with a RDF knowledge base (YAGO2s) is known as XKG (eXtended Knowledge Graph). The triple scores for YAGO2s triples are equal to the number of inlinks into the subject, i.e., the number of times the entity in the subject occurs in the object of any triple. XKG has about $105$ million triples. This dataset was selected for having a rich variety of relaxations.
  We evaluated on $65$ queries which were manually constructed so as to have non-empty result sets. Each query has $2$-$4$ triple patterns and each triple pattern has atleast $10$ relaxations. The relaxations were obtained using the scheme outlined in \cite{xkg2016}. 

 \item Twitter tweets:\\
  The dataset was built using Twitter Streaming API over trending hashtags. The stream was tracked for $30$ days with each day's trending tags over the month of April $2017$. The dataset has about $18$ million unique triples of the form: $\langle tID,hasTag,T \rangle$ where $tID$ is the unique ID for a tweet and $T$ is a term contained in the tweet with ID as $tID$. A query over this dataset queries for IDs of those tweets which have all the queried terms. For example, the following query queries for IDs of all those tweets which contain the terms `\#intoyouvideo', `\#ariana' and `dangerous':

  SELECT ?s WHERE\{\\
  \texttt{
  \hspace*{5mm}?s <hasTag> <\#intoyouvideo>.\\
  \hspace*{5mm}?s <hasTag> <\#ariana>.\\
  \hspace*{5mm}?s <hasTag> <dangerous>}\\
  \}
  
  The score for each triple is equal to the number of retweets for the tweet in that triple. The relaxations were generated using the co-occurrence frequencies i.e. the relaxation weight, $w$ for the relaxation, $r=(T_1,T_2,w)$ will be equal to:
  $$ w=\frac{\#tweets\_having\_T_1\_and\_T_2}{\#tweets\_having\_T_1}$$

  For example, a possible relaxation for \term{<\#intoyouvideo>} is \term{<video>}.

  The testset of $50$ queries was constructed manually using combinations of most frequent tags and terms. Each query had either $2$ or $3$ triple patterns, with each triple pattern having atleast $5$ relaxations. 
\end{enumerate}

\subsection{Metrics}
We measure the following metrics for each query to demonstrate the quality and efficiency of our technique:
\begin{enumerate}
 \item Quality:
 \begin{itemize}
  \item Precision: The fraction of true top-$k$ results (of TriniT) in the top-$k$ results of Spec-QP.
  \item Recall: The fraction of top-$k$ results by Spec-QP in the true top-$k$ by TriniT.
  \item Prediction accuracy: The number of queries for which we could identify the correct relaxations.
  \item Score error: The average of absolute error for Spec-QP vs. TriniT top-$k$ scores, i.e., \\ $\frac{1}{k}\sum_{i=1..k}\abs{score_i^{Spec-QP}-score_i^{TriniT}}$\\
  We also note the standard deviation.
 \end{itemize}
 \item Efficiency:
 \begin{itemize}
  \item Runtimes: We measure the time taken to plan and execute each query.
  \item Memory used: Since it is not easy to measure exact memory consumption in Java, we use no. of answer objects created as a means to represent this. The total no. of answer objects created directly corresponds to the amount of search space traversed to arrive at top-$k$ answers. This number includes all the intermediate answer objects encountered by Incremental Merges and Rank Joins.
 \end{itemize}
\end{enumerate}
Note that precision and recall have identical values in our setup, because they have the same denominator $k$.

\subsection{Setup}
The experiments were conducted on a Dell Blade server with $24$ Intel(R) Xeon(R) CPU E5-2420 @ 1.90GHz processors and $32$GB RAM. The database engine used to retrieve the matches for triple patterns in sorted order is postgresql-9.5. Each query was evaluated using both the techniques- TriniT and Spec-QP. We considered three values for $k$, namely $10$, $15$ and $20$. To have a warm cache, we conducted $5$ consecutive runs for each query and considered the average of the last $3$ runs for each technique.

\subsection{Quality evaluations}
We first discuss the quality of results obtained by Spec-QP and then provide the statistics for runtimes and memory consumptions.
\begin{table}
\centering
\begin{tabular}{ |p{0.5cm}|p{2cm}|p{2cm}|}
\hline
 \textbf{k} & \textbf{XKG} & \textbf{Twitter}\\ 
 \hline 
  10&0.7&0.72\\
 \hline
  15&0.88&0.78\\
 \hline
  20&0.91&0.8\\
 \hline
\end{tabular}
\caption{Precision (and Recall) over each dataset.}\label{tab:prec}
\end{table}
  
\subsubsection{Precision (and Recall)}

The precision values for the datasets are given in Table \ref{tab:prec}. The precision is good for both the datasets, being about 90\% in the best case. This indicates that on an average $90$\% of the answers belonged to true top-$k$. Note that since the answers are sorted according to the scores, the answers outside the true top-$k$ appeared at lower ranks. The higher ranked answers were found correctly. An observation is that the accuracy increased with increasing the value for $k$. Also, the approximate results obtained are very close to true top-$k$ as described later in Section \ref{sec:avg-err}.

\subsubsection{Prediction Accuracy}
We performed a detailed analysis of the number of queries for which we could predict the correct relaxation(s) over each dataset. They are given in Table \ref{tab:all-prec-results}. We observed that each query required some triple patterns to be relaxed to generate top-$k$ answers. It can be seen that the prediction accuracy is atleast $70$\% for all types of queries over XKG and queries requiring $3$ relaxations over Twitter. As the value for $k$ was increased, queries increasingly required relaxations to generate sufficient answers. For twitter, most of the queries required all triple patterns to be relaxed. This is due to absence of sufficient triples corresponding to each term and fewer relaxations (predicate does not have relaxations) for each triple pattern. Nevertheless, we were able to identify the requirement of all the relaxations in such a scenario. 

\begin{table*}[ht!]
\centering
\begin{tabular}{ |l|l|l|l||l|l|l|}
\hline
Dataset&
  \multicolumn{3}{c||}{XKG} &
  \multicolumn{3}{c|}{Twitter} \\
  \hline
  
   {k} & {10} &{15}  & {20} &{10} &{15}  & {20} \\

 \hline 
 queries requiring $1$ relaxation & 5(6) & 5(5) & -(-) & - & - & - \\
 \hline
 queries requiring $2$ relaxations & 21(30) & 22(26) & 18(19) & 1(2) & 1(2) & 1(2)  \\
 \hline
 queries requiring $3$ relaxations & 12(18) & 16(19) & 27(31) & 35(48) & 38(48) & 39(48) \\
 \hline
 queries requiring $4$ relaxations & 7(11) & 14(15) & 14(15) & - & - & - \\
 \hline
 \end{tabular}
\caption{Summary of prediction accuracy for various values of $k$ over XKG and Twitter grouped by the number of triple patterns requiring relaxations in the queries to generate true top-$k$ results. The number indicates the number of queries for which Spec-QP could identify exactly only these relaxations. The numbers in brackets show the total number of such queries.}\label{tab:all-prec-results}
\end{table*}

Note that the $2$-bucket histogram model for representing the distribution of scores is only an estimation of the score distribution and not the exact distribution. This leads to wrong estimates for expected score values in few cases. This can be improved upon by using multi-bucket histograms for modelling exact distributions but it will lead to higher planning time overheads.

\subsubsection{Average score error}\label{sec:avg-err}
To judge the quality of approximate results returned by Spec-QP, we computed the score deviations of the approximate answers at each rank given by Spec-QP from the true top-$k$. The average values for the score difference (along with the standard deviations) for various values of $k$ are given in Table \ref{tab:rank-dev}. The percentages in brackets show the average percentage deviation from the original scores. Note that the maximum possible score for an answer to a $2$ triple pattern query can be $2$, for a $3$ triple pattern query, it will be $3$ and so on.\footnote{This is because the maximum score for a matching triple for each triple pattern can be $1$.}

\begin{table*}[ht!]
\centering
\begin{tabular}{ |l|l|l|l||l|l|}
\hline
Dataset&
  \multicolumn{3}{c||}{XKG} &
  \multicolumn{2}{c|}{Twitter} \\ 
  \hline 
\backslashbox{k}{\#TP}& {2} &{3}  & {4} &{2} &{3}\\
 \hline 
  10&0.1(5\%)$\pm$0.1&0.2(8\%)$\pm$0.3&0.1(3\%)$\pm$0.2&0.16(8\%)$\pm$0.0&0.5(16\%)$\pm$0.5\\
 \hline
  15&0.08(4\%)$\pm$0.08&0.1(3\%)$\pm$0.2&0.01(1\%)$\pm$0.04&0.16(8\%)$\pm$0.0&0.32(10\%)$\pm$0.3\\
 \hline
  20&0.07(4\%)$\pm$0.06&0.07(2\%)$\pm$0.1&0.01(1\%)$\pm$0.03&0.16(8\%)$\pm$0.0&0.18(6\%)$\pm$0.1\\
 \hline
\end{tabular}
\caption{Average score deviations for the approximate top-$k$ from the true top-$k$ for each dataset. It is grouped by the number of triple patterns (\#TP) in the queries. The percentages in brackets show the average percentage deviation from the original scores.}\label{tab:rank-dev}
\end{table*}
\paragraph*{\textbf{XKG}} Even though k=$10$ has lowest precision, the score deviations from true top-$k$ answers are low (about $0.1$ for $2$ triple pattern queries). That is, for a query with $2$ triple patterns if the actual answer at a given rank has a score of $1.5$, the score of the approximate answer would be about $1.4$. The deviations are even lower (only about $0.01$) for higher values of $k$ and tolerable for achieving faster runtimes. In agreement with the trend for precision values, the deviation reduces as we increase $k$.

\paragraph*{\textbf{Twitter}} There is only one $2$ triple pattern query that required both the triple patterns to be relaxed but had a wrong speculation of relaxations for all values of $k$. However, its score deviation is constant over all values of $k$ due to it having only $11$ results (including relaxations). The deviations are only $0.5$ for $3$ triple pattern queries with $k=10$, which is only $16$\% deviation from the original scores. The deviations for higher values of $k$ are very low being only $6$\% in the best case. For k=$20$, for a query with $3$ triple patterns if the actual answer at a given rank has a score of $2.5$, the score of the approximate answer would be about $2.32$.

\paragraph*{Summary of precision results}
We showed that our predictor predicted the correct relaxations about $70$-$80$\% of the time as can be seen from the precision analysis. Also, the answers outside the top-$k$ had minimal score deviations from the original top-$k$ answers at each rank (Table \ref{tab:rank-dev}) indicating that Spec-QP misses the true top-$k$ only narrowly. Hence, Spec-QP gives approximate top-$k$ of good quality.

\subsection{Efficiency evaluations}
We now discuss the efficiency of Spec-QP over TriniT in generating the results for individual datasets.

\subsubsection{Efficiency over XKG}

\begin{figure}[!ht]

\begin{subfigure}[b]{0.3\textwidth}
\includegraphics[width=\textwidth]{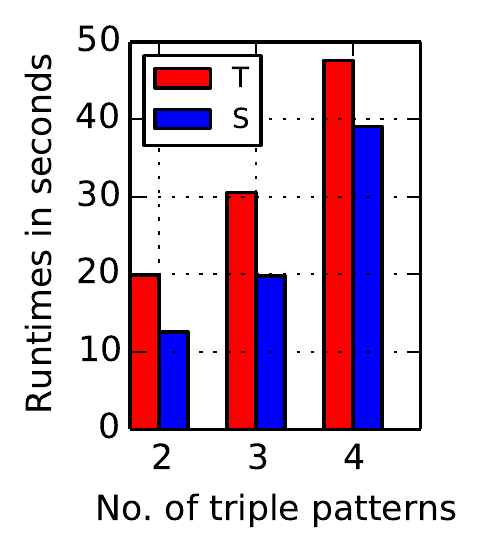}\\
        \caption{Runtimes for k=$10$. }
        \label{fig:xkg-k10r}
\end{subfigure}
\begin{subfigure}[b]{0.3\textwidth}.
\includegraphics[width=\textwidth]{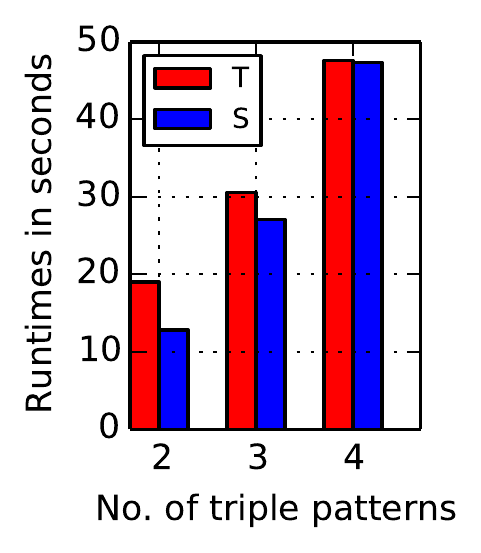}\\
        \caption{Runtimes for k=$15$. }
        \label{fig:xkg-k15r}
\end{subfigure}
\begin{subfigure}[b]{0.3\textwidth}
\includegraphics[width=\textwidth]{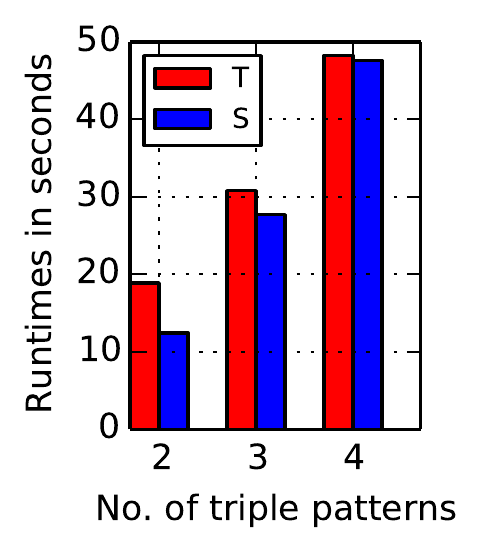}\\
        \caption{Runtimes for k=$20$. }
        \label{fig:xkg-k20r}
\end{subfigure}
\\
\begin{subfigure}[b]{0.3\textwidth}
 \includegraphics[width=\textwidth]{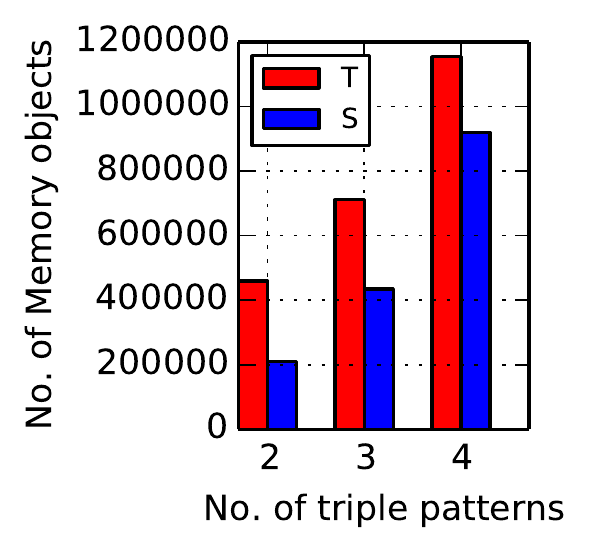}
        \caption{Memory for k=$10$.}
        \label{fig:xkg-k10m}
\end{subfigure}
\begin{subfigure}[b]{0.3\textwidth}
 \includegraphics[width=\textwidth]{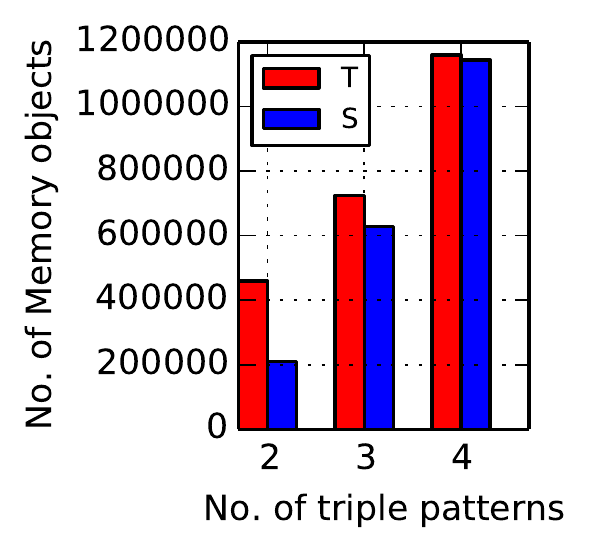}
        \caption{Memory for k=$15$.}
        \label{fig:xkg-k15m}
\end{subfigure}
\begin{subfigure}[b]{0.3\textwidth}
 \includegraphics[width=\textwidth]{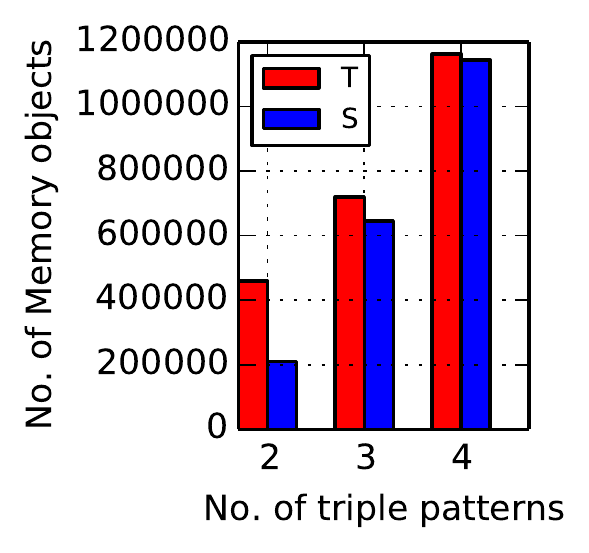}
        \caption{Memory for k=$20$.}
        \label{fig:xkg-k20m}
\end{subfigure}
\caption{Runtimes and memory comparisons over XKG queries for k=$10$, $15$ and $20$ grouped by the no. of triple patterns in the query. \textbf{All the legends in the graphs for efficiency have `T' for TriniT and `S' for Spec-QP.}}
\label{fig:xkg-results-numtp}
\end{figure}
The results for XKG grouped by the number of triple patterns in the queries have been given in Figure \ref{fig:xkg-results-numtp}.
\begin{itemize}
 \item k=$10$: Spec-QP outperforms TriniT by a great margin for $k=10$. This is because Spec-QP avoids unnecessary computation of \textit{all} relaxations when only few relaxations are capable of giving top-$k$ answers. Most of the queries require only $2$ relaxations (Refer Table \ref{tab:all-prec-results}) to produce top-$10$ answers and Spec-QP either identifies the correct relaxation(s) or gives good quality approximate results.

 \item k=$15$ and k=$20$: Here, the $2$ and $3$ triple pattern queries have faster runtimes on using Spec-QP. The gain margin however has lowered from previous value of $k$. This is because when the user seeks more answers, the original query becomes increasingly insufficient in generating $k$ answers and more relaxations are required. For $4$ triple pattern queries, higher values of $k$ leads to more relaxations because of answers becoming sparse with each join. Hence, the runtimes and memory consumptions are closer to TriniT.

\end{itemize}

\begin{figure}[!ht]

\begin{subfigure}[b]{0.3\textwidth}
\includegraphics[width=\textwidth]{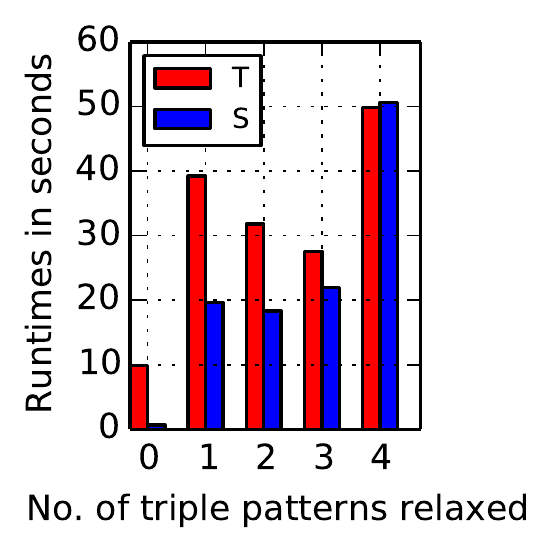}\\
        \caption{Runtimes for k=$10$. }
        \label{fig:xkg-k10r-nr}
\end{subfigure}
\begin{subfigure}[b]{0.3\textwidth}
\includegraphics[width=\textwidth]{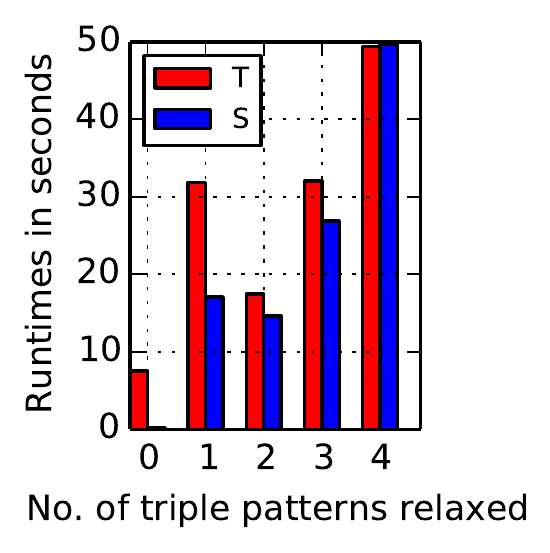}\\
        \caption{Runtimes for k=$15$. }
        \label{fig:xkg-k15r-nr}
\end{subfigure}
\begin{subfigure}[b]{0.3\textwidth}
\includegraphics[width=\textwidth]{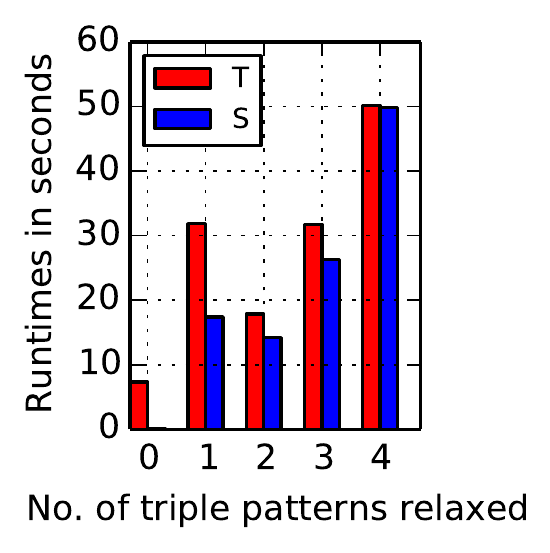}\\
        \caption{Runtimes for k=$20$. }
        \label{fig:xkg-k20r-nr}
\end{subfigure}
\\

\begin{subfigure}[b]{0.3\textwidth}
 \includegraphics[width=\textwidth]{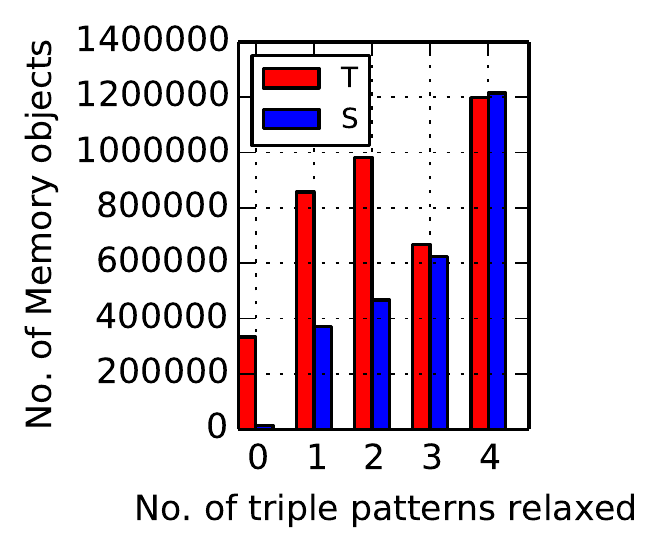}
        \caption{Memory for k=$10$.}
        \label{fig:xkg-k10m-nr}
\end{subfigure}
\begin{subfigure}[b]{0.3\textwidth}
 \includegraphics[width=\textwidth]{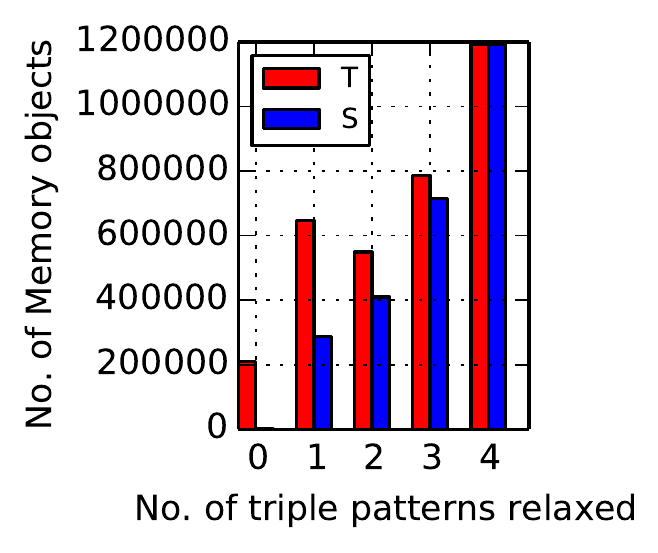}
        \caption{Memory for k=$15$.}
        \label{fig:xkg-k15m-nr}
\end{subfigure}
\begin{subfigure}[b]{0.3\textwidth}
 \includegraphics[width=\textwidth]{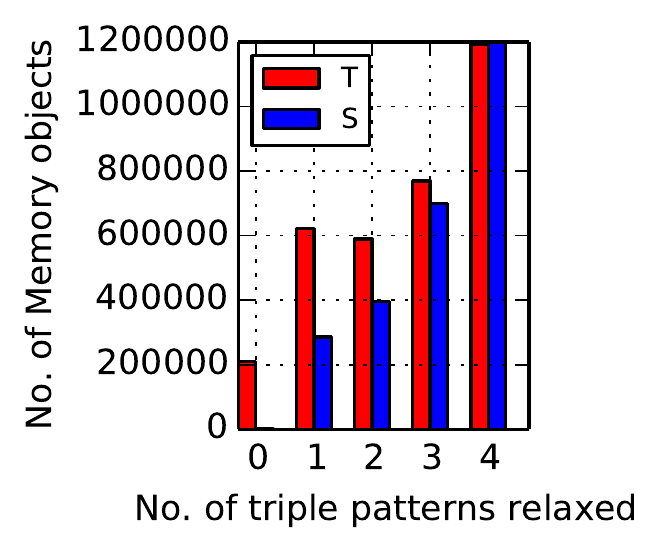}
        \caption{Memory for k=$20$.}
        \label{fig:xkg-k20m-nr}
\end{subfigure}
\caption{Runtimes and memory comparisons over XKG queries for k=$10$, $15$ and $20$ grouped by the no. of triple patterns relaxed in the query by Spec-QP. \textbf{All the legends in the graphs for efficiency have `T' for TriniT and `S' for Spec-QP.}}
\label{fig:xkg-results-nr}
\end{figure}

The results grouped by the number of triple patterns relaxed by Spec-QP in the queries for XKG have been given in Figure \ref{fig:xkg-results-nr}. We can see that we have major gains when none of the triple patterns undergo relaxations. The difference in the runtimes of TriniT and Spec-QP reduces when more no. of triple patterns are relaxed. This is because with increasing no. of triple patterns requiring relaxations, the Spec-QP plan tends towards the plan by TriniT, i.e., processing relaxations from all the triple patterns. The memory consumption also follows a similar trend. For cases with $4$ triple patterns relaxed, all the triple patterns in the query are relaxed. The runtimes in these cases are slightly higher than TriniT owing to the additional time spent on speculative planning. The memory consumption is the same as for TriniT.

\subsubsection{Efficiency over Twitter}
\begin{figure}[!ht]

\begin{subfigure}[b]{0.3\textwidth}
\includegraphics[width=\textwidth]{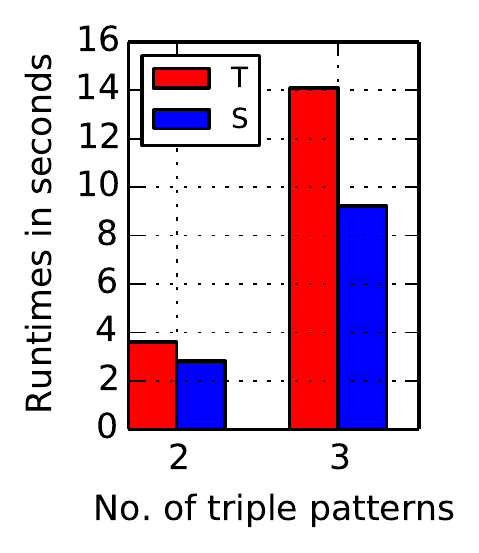}\\
        \caption{Runtimes for k=$10$. }
        \label{fig:twitter-k10r}
\end{subfigure}
\begin{subfigure}[b]{0.3\textwidth}
\includegraphics[width=\textwidth]{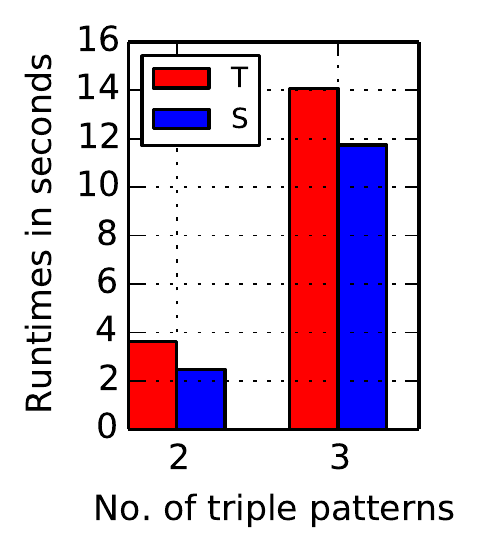}\\
        \caption{Runtimes for k=$15$. }
        \label{fig:twitter-k15r}
\end{subfigure}
\begin{subfigure}[b]{0.3\textwidth}
\includegraphics[width=\textwidth]{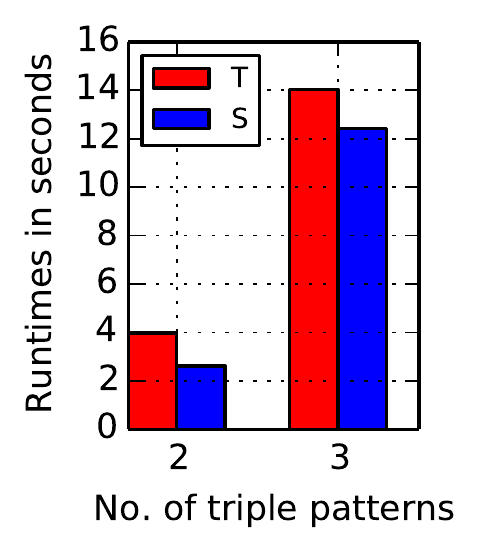}\\
        \caption{Runtimes for k=$20$. }
        \label{fig:twitter-k20r}
\end{subfigure}
\\

\begin{subfigure}[b]{0.3\textwidth}
 \includegraphics[width=\textwidth]{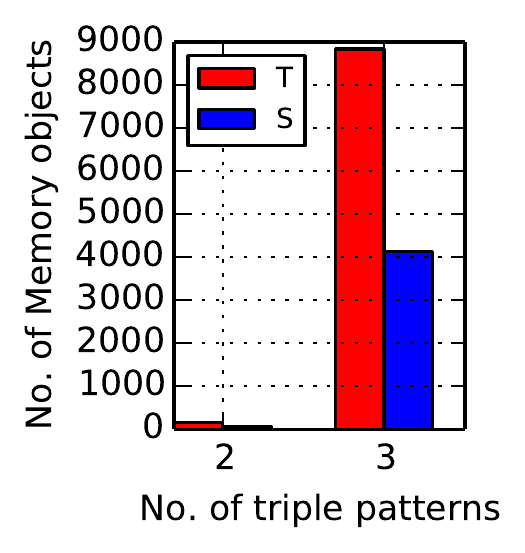}
        \caption{Memory for k=$10$.}
        \label{fig:twitter-k10m}
\end{subfigure}
\begin{subfigure}[b]{0.3\textwidth}
 \includegraphics[width=\textwidth]{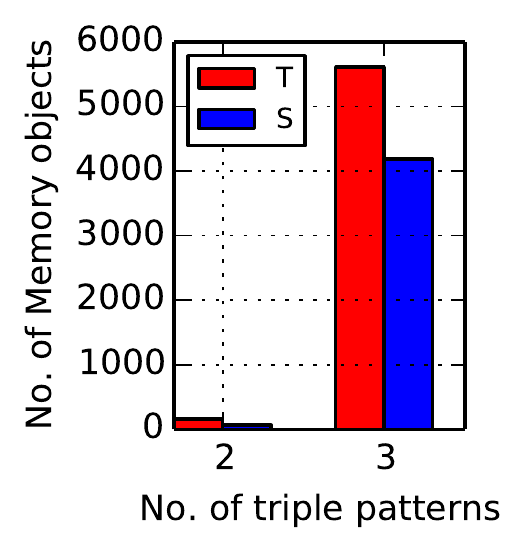}
        \caption{Memory for k=$15$.}
        \label{fig:twitter-k15m}
\end{subfigure}
\begin{subfigure}[b]{0.3\textwidth}
 \includegraphics[width=\textwidth]{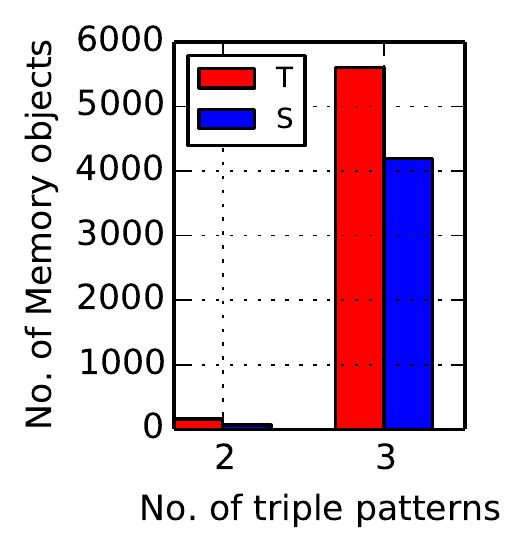}
        \caption{Memory for k=$20$.}
        \label{fig:twitter-k20m}
\end{subfigure}
\caption{Runtimes and memory comparisons over Twitter for k=$10$, $15$ and $20$ grouped by the no. of triple patterns in the query. \textbf{All the legends in the graphs for efficiency have `T' for TriniT and `S' for Spec-QP.}}
\label{fig:twitter-results}
\end{figure}

The results grouped by the number of triple patterns in the queries over Twitter data for various values of $k$ are given in Figure \ref{fig:twitter-results}. 
\begin{itemize}
 \item k=$10$: We can see here that the Spec-QP performs really well on all queries. We have faster response times. The memory used is also less than the TriniT plan. As discussed before, Spec-QP identifies the required relaxations ascertaining low score deviations from true top-$k$ for all the queries.
 \item k=$15$ and $20$: The results are similar to what was observed for k=$10$. Additionally, we observe that with increasing value of $k$, the difference between the runtimes of Spec-QP and TriniT reduces from that for k=$10$. This is due to the fact that when the user demands more answers, the original query no longer has sufficient answers. Hence, triple patterns require relaxations and the query requires more time to execute.
\end{itemize}

\begin{figure}[!ht]

\begin{subfigure}[b]{0.3\textwidth}
\includegraphics[width=\textwidth]{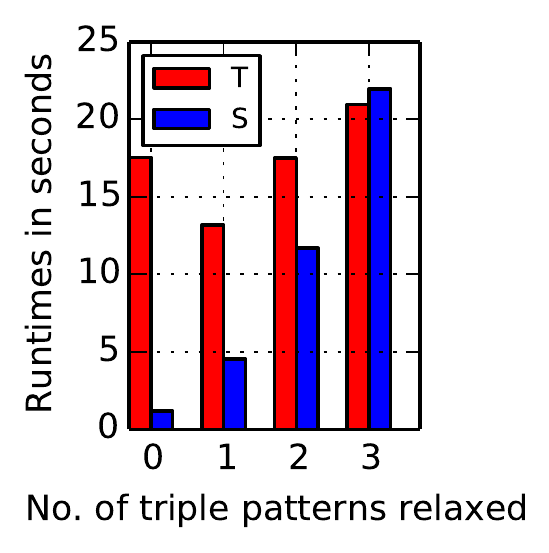}\\
        \caption{Runtimes for k=$10$. }
        \label{fig:twitter-k10r-nr}
\end{subfigure}
\begin{subfigure}[b]{0.3\textwidth}
\includegraphics[width=\textwidth]{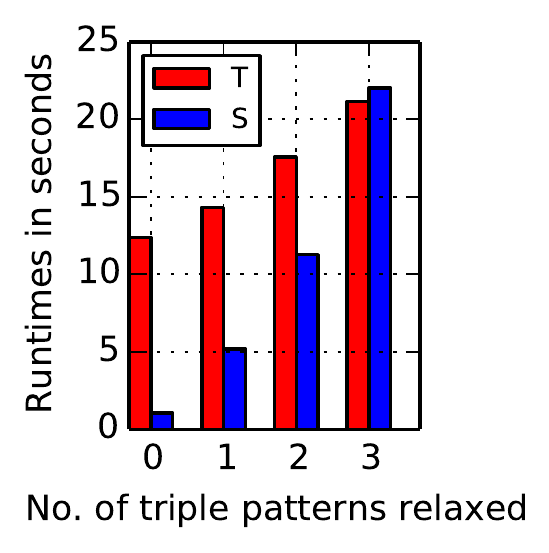}\\
        \caption{Runtimes for k=$15$. }
        \label{fig:twitter-k15r-nr}
\end{subfigure}
\begin{subfigure}[b]{0.3\textwidth}
\includegraphics[width=\textwidth]{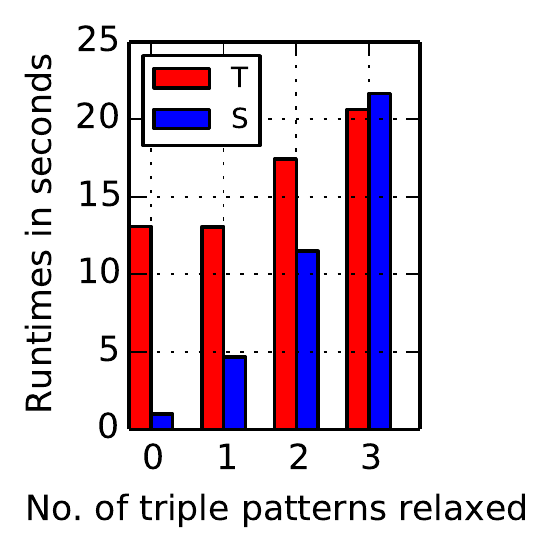}\\
        \caption{Runtimes for k=$20$. }
        \label{fig:twitter-k20r-nr}
\end{subfigure}
\\

\begin{subfigure}[b]{0.3\textwidth}
 \includegraphics[width=\textwidth]{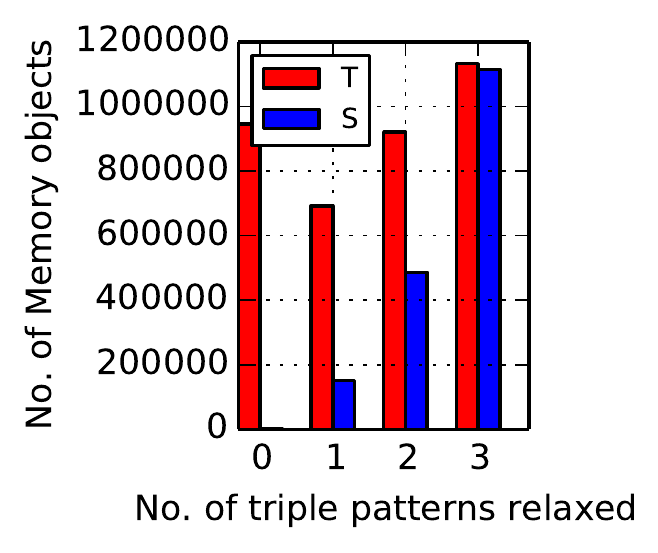}
        \caption{Memory for k=$10$.}
        \label{fig:twitter-k10m-nr}
\end{subfigure}
\begin{subfigure}[b]{0.3\textwidth}
 \includegraphics[width=\textwidth]{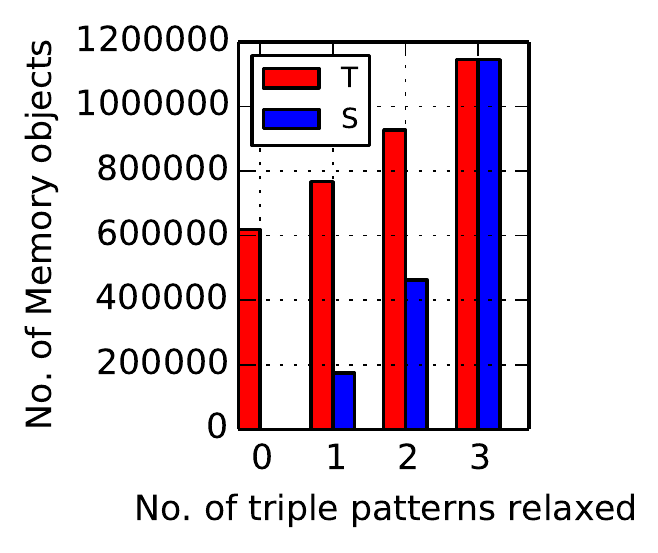}
        \caption{Memory for k=$15$.}
        \label{fig:twitter-k15m-nr}
\end{subfigure}
\begin{subfigure}[b]{0.3\textwidth}
 \includegraphics[width=\textwidth]{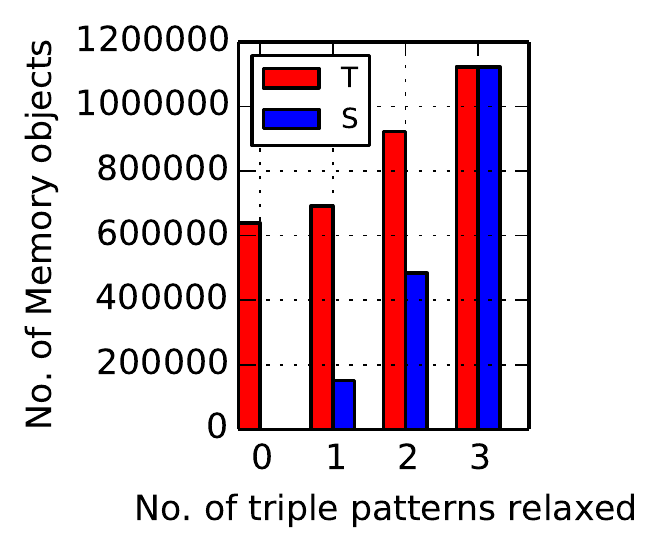}
        \caption{Memory for k=$20$.}
        \label{fig:twitter-k20m-nr}
\end{subfigure}
\caption{Runtimes and memory comparisons over Twitter for k=$10$, $15$ and $20$ grouped by the no. of triple patterns relaxed in the query by Spec-QP. \textbf{All the legends in the graphs for efficiency have `T' for TriniT and `S' for Spec-QP.}}
\label{fig:twitter-results-nr}
\end{figure}

The results grouped by the number of triple patterns relaxed by Spec-QP in the queries over Twitter data for various values of $k$ are given in Figure \ref{fig:twitter-results-nr}. The results are similar to what was observed for XKG. For queries requiring $3$ relaxations, Spec-QP is similar to TriniT since relaxations from all the triple patterns are processed. The runtimes in these cases are slightly higher than TriniT owing to the additional time spent on speculative planning. The memory consumption is the same as for TriniT.

\subsection{Discussion and remarks}
We have shown that Spec-QP is able to identify the correct relaxation(s) for most of the queries. For queries having precision $<1$, Spec-QP gives good quality approximations for top-$k$, as is demonstrated by the average score deviation values. We have also shown that Spec-QP incurs less computation overheads and achieves faster response times with low memory overheads.  Hence, Spec-QP is more efficient than TriniT and also has good accuracy.

\section{Related Work}\label{sec:related}

\subsection*{Top-$k$ query processing}
 FRPA \cite{frpa-rankjoin} and Hash Rank-Join (HRJN*) \cite{rankjoin2004} represent the state-of-the-art relational rank-join algorithms. HRJN* has been shown to perform well in practice, however, FRPA showed that it was not instance-optimal for a variant of the rank join problem that they considered. HRJN\cite{ilyas@sigmod2004} is based on ripple join algorithm. It maintains two hash tables in-memory for storing the input tuples seen so far, the stored input tuples are used for finding join results. These results are then given as inputs to a priority queue, which outputs them in the order specified by the ranking function. Nested Loops Rank Join (NRJN) \cite{rankjoin2003} is similar to HRJN except that unlike HRJN it does not store input tuples, but rather follows a nested-loop strategy.  Pull/Bound Rank Join (PBRJ) \cite{pbrj} is an algorithm template that generalized previous rank join algorithms and provided tight upper bounds. DRJN \cite{rankjoindist} is an efficient algorithm for computing rank joins in distributed systems. Theobald et. al. \cite{theobald@vldb2004} dealt with top-k query evaluation for joins over multiple index lists with pruning using probabilistic guarantees. It does so using histograms and dynamic convolutions to predict the top-$k$. Our case, however differs in that we consider graph structured data and also, support multiple relaxations. The IO-Top-k \cite{io-top-k} deals with top-$k$ query evaluation with pruning using sorted access (SA) scheduling. Other works include top-$k$ processing over xml data \cite{top-x} and for data that is distributed over multiple nodes \cite{yu@dexa2005}. 

\paragraph*{Top-$k$ queries on graphs}
There are very few works which address the problem of top-$k$ processing over RDF graphs. The SPARQL-RANK framework proposed by \cite{magliacane@iswc2012} makes use of different index permutations used in native triple-stores for fast random access during top-k processing, and applies early-termination criterion. They propose an algorithm, which requires the left-most index used in the join plan to be sorted based on the ranking function, and then it randomly probes the right-side index. Thus, when the right-side index is large, the performance of rank join suffers. In another framework introduced by Wang et al. \cite{wang@IS2015}, quantitative entities in the RDF dataset are separated out into an MS-tree index. In the first step of query processing, candidate entities are located using the MS-tree index that are then used as seeds for performing breadth-first (BFS) traversals over the graph to find matching sub-graphs. If the query requires only a few highly correlated predicates, the algorithm may end up storing many unnecessary nodes in the queue, making the retrieval of the first entity possible only after several iterations. The work in \cite{yang@icde2016} uses an approach similar to HRJN\cite{ilyas@sigmod2004} for computing top-k star joins. However for RDF data, SPARQL-RANK showed experimentally that it outperformed HRJN. The performance gain was attributed to the unsorted nature of numerical attributes present in indexes build by RDF engines. QUARK-X \cite{quark-x} proposes an efficient technique to process top-k queries on RDF graphs using extra indexes and metadata. Another work specific to Linked Data is by Wagner et. al. \cite{wagner@eswc2012}, where partial results are located at different sources and can only be accessed via URI lookups. All of these works however do not consider efficient processing for relaxations over the original query. 

\subsection*{Query Reformulation in IR}
Various strategies have been proposed to reformulate queries in IR over documents. These include measures of query similarity \cite{baeza-yates@edbt2004}, or using summary information included in the query-flow graph \cite{anagnostopoulos@wsdm10}. Another approach by Hristidis et. al. \cite{hristidis@2010} relies on suggesting keyword relaxations by relaxing those which are least specific based on their idf score. These reformulations can be used as relaxations for our setting.

\subsection*{Faceted Search: Many answers problem}
A related optimization problem is the one encountered when we have many-answers, i.e. those where given an initial query that returns a large number of answers, the
objective is to design an effective drill-down strategy to help the user find acceptable results with minimum effort \cite{roy@cikm2008, kashyap@cikm2010, li@www2010}. We solve a related problem, where we try to solve both empty-answer and many-answers problem in an efficient manner by generating additional scored answers using relaxations.

\subsection*{Query Relaxation Frameworks}
Query relaxation in relational databases is quite common. The work \cite{koudas@2006} relaxes joins and selections in relational databases by suggesting alternative queries based on the ``minimal'' shift from the original query. Another work \cite{vasilyeva@adbis2014} suggests user ranking of the query edges so as to generate relevant differential queries with minimum deviation. ``Why Not'' queries are studied in \cite{chapman@sigmod2009, tran@sigmod2010}, where, given a query Q that did not return a set of tuples S that the user was expecting to be returned, they design an alternate query Q' that (a) is very similar to Q, and (b) returns the missing tuples S, however the rest of the returned tuples should not be too different from those returned by Q. The paper \cite{prob-empty-answer} relaxes one constraint at a time and is interactive. It also tries to minimize the cost by suggesting low cost relaxations which lead to non-empty answers. DebEAQ \cite{DebEAQ} first tries to debug why the query is returning empty answer and then tries to relax it with minimum change to the original query. It is also limited only to property graphs.

The closest to our works are those which deal with relaxations over graphs. The paper \cite{poulovassilis@iswc2010} considers query relaxation for conjunctive regular path queries. Users are able to specify approximations and relaxations to be applied to their original query, and the relative costs of these. Query results are returned incrementally, ranked in order of increasing distance from the user’s original query. Another work which computes approximates answers uses two algorithms for evaluation \cite{huang@www2012}. The first algorithm is based on best-first strategy and relaxed queries are executed in order. They prune relaxations which do not give new results. The other algorithm executes the relaxed queries as a batch and avoids the unnecessary execution cost. TriniT \cite{xkg2016} enhances the graphs using text corpus and computes relaxations over them. The relaxations are computed efficiently using incremental merges and rank joins. We use this system as our baseline.

\section{Conclusion and Future Work}\label{sec:conclusion}
We have proposed Spec-QP, a strategy for top-$k$ query processing in a scenario where a query can have multiple relaxations. To achieve this, we used a speculative approach for pruning the relaxations which are not likely to contribute answers to the top-$k$ results. The triple patterns which are predicted to not require relaxations can be processed by rank joins over the sorted list of matches for them, thereby reducing top-$k$ processing and leading to great savings on runtimes and memory. Extensive experiments over two real world datasets - XKG and Twitter, show that Spec-QP achieves greater efficiency over the baseline with good accuracy for most of the queries. As future work, we would like to generate and use more complicated relaxations for the queries like replacing a triple pattern with a chain of triple patterns, etc. Also, we would like to extend these techniques to work for ranked retrieval from XML databases.


\begin{thebibliography}{10}

\bibitem{anagnostopoulos@wsdm10}
Aris Anagnostopoulos, Luca Becchetti, Carlos Castillo, and Aristides Gionis.
\newblock An optimization framework for query recommendation.
\newblock In {\em Proceedings of the Third International Conference on Web
  Search and Web Data Mining, {WSDM} 2010, New York, NY, USA, February 4-6,
  2010}, 2010.

\bibitem{dbpedia}
S{\"{o}}ren Auer, Christian Bizer, Georgi Kobilarov, Jens Lehmann, Richard
  Cyganiak, and Zachary~G. Ives.
\newblock Dbpedia: {A} nucleus for a web of open data.
\newblock In {\em The Semantic Web, 6th International Semantic Web Conference,
  2nd Asian Semantic Web Conference, {ISWC} 2007 + {ASWC} 2007, Busan, Korea,
  November 11-15, 2007.}, 2007.

\bibitem{baeza-yates@edbt2004}
Ricardo~A. Baeza{-}Yates, Carlos~A. Hurtado, and Marcelo Mendoza.
\newblock Query recommendation using query logs in search engines.
\newblock In {\em Current Trends in Database Technology - {EDBT} 2004
  Workshops, {EDBT} 2004 Workshops PhD, DataX, PIM, P2P{\&}DB, and ClustWeb,
  Heraklion, Crete, Greece, March 14-18, 2004, Revised Selected Papers}, 2004.

\bibitem{io-top-k}
H.~Bast, Debapriyo Majumdar, Ralf Schenkel, Martin Theobald, and Gerhard
  Weikum.
\newblock Io-top-k: Index-access optimized top-k query processing.
\newblock In {\em Proceedings of the 32nd International Conference on Very
  Large Data Bases, Seoul, Korea, September 12-15, 2006}, 2006.

\bibitem{freebase}
Kurt~D. Bollacker, Robert~P. Cook, and Patrick Tufts.
\newblock Freebase: {A} shared database of structured general human knowledge.
\newblock In {\em Proceedings of the Twenty-Second {AAAI} Conference on
  Artificial Intelligence, July 22-26, 2007, Vancouver, British Columbia,
  Canada}, 2007.

\bibitem{chapman@sigmod2009}
Adriane Chapman and H.~V. Jagadish.
\newblock Why not?
\newblock In {\em Proceedings of the {ACM} {SIGMOD} International Conference on
  Management of Data, {SIGMOD} 2009, Providence, Rhode Island, USA, June 29 -
  July 2, 2009}, 2009.

\bibitem{david2004order}
H.A. David and H.N. Nagaraja.
\newblock {\em Order Statistics}.
\newblock Wiley Series in Probability and Statistics. Wiley, 2004.

\bibitem{rankjoindist}
Christos Doulkeridis, Akrivi Vlachou, Kjetil N{\o}rv{\aa}g, Yannis Kotidis, and
  Neoklis Polyzotis.
\newblock Processing of rank joins in highly distributed systems.
\newblock In {\em {IEEE} 28th International Conference on Data Engineering
  {(ICDE} 2012), Washington, DC, {USA} (Arlington, Virginia), 1-5 April, 2012},
  2012.

\bibitem{elbassuoni@cikm2009}
Shady Elbassuoni, Maya Ramanath, Ralf Schenkel, Marcin Sydow, and Gerhard
  Weikum.
\newblock Language-model-based ranking for queries on rdf-graphs.
\newblock In {\em Proceedings of the 18th {ACM} Conference on Information and
  Knowledge Management, {CIKM} 2009, Hong Kong, China, November 2-6, 2009},
  2009.

\bibitem{query-relax}
Shady Elbassuoni, Maya Ramanath, and Gerhard Weikum.
\newblock Query relaxation for entity-relationship search.
\newblock In {\em The Semanic Web: Research and Applications - 8th Extended
  Semantic Web Conference, {ESWC} 2011, Heraklion, Crete, Greece, May 29 - June
  2, 2011, Proceedings, Part {II}}, 2011.

\bibitem{colina}
Azam Feyznia, Mohsen Kahani, and Fattane Zarrinkalam.
\newblock {COLINA:} {A} method for ranking {SPARQL} query results through
  content and link analysis.
\newblock In {\em Proceedings of the {ISWC} 2014 Posters {\&} Demonstrations
  Track a track within the 13th International Semantic Web Conference, {ISWC}
  2014, Riva del Garda, Italy, October 21, 2014.}, 2014.

\bibitem{frpa-rankjoin}
Jonathan Finger and Neoklis Polyzotis.
\newblock Robust and efficient algorithms for rank join evaluation.
\newblock In {\em Proceedings of the {ACM} {SIGMOD} International Conference on
  Management of Data, {SIGMOD} 2009, Providence, Rhode Island, USA, June 29 -
  July 2, 2009}.

\bibitem{hristidis@2010}
Vagelis Hristidis, Yuheng Hu, and Panagiotis~G. Ipeirotis.
\newblock Ranked queries over sources with boolean query interfaces without
  ranking support.
\newblock In {\em Proceedings of the 26th International Conference on Data
  Engineering, {ICDE} 2010, March 1-6, 2010, Long Beach, California, {USA}},
  2010.

\bibitem{huang@www2012}
Hai Huang, Chengfei Liu, and Xiaofang Zhou.
\newblock Approximating query answering on {RDF} databases.
\newblock {\em World Wide Web}, 15(1), 2012.

\bibitem{rankjoin2003}
Ihab~F. Ilyas, Walid~G. Aref, and Ahmed~K. Elmagarmid.
\newblock Supporting top-k join queries in relational databases.
\newblock In {\em {VLDB}}, 2003.

\bibitem{rankjoin2004}
Ihab~F. Ilyas, Walid~G. Aref, and Ahmed~K. Elmagarmid.
\newblock Supporting top-k join queries in relational databases.
\newblock {\em {VLDB} J.}, 13(3), 2004.

\bibitem{ilyas@sigmod2004}
Ihab~F. Ilyas, Rahul Shah, Walid~G. Aref, Jeffrey~Scott Vitter, and Ahmed~K.
  Elmagarmid.
\newblock Rank-aware query optimization.
\newblock In {\em Proceedings of the {ACM} {SIGMOD} International Conference on
  Management of Data, Paris, France, June 13-18, 2004}, 2004.

\bibitem{kashyap@cikm2010}
Abhijith Kashyap, Vagelis Hristidis, and Michalis Petropoulos.
\newblock Facetor: cost-driven exploration of faceted query results.
\newblock In {\em Proceedings of the 19th {ACM} Conference on Information and
  Knowledge Management, {CIKM} 2010, Toronto, Ontario, Canada, October 26-30,
  2010}, 2010.

\bibitem{naga}
Gjergji Kasneci, Fabian~M. Suchanek, Georgiana Ifrim, Maya Ramanath, and
  Gerhard Weikum.
\newblock {NAGA:} searching and ranking knowledge.
\newblock In {\em Proceedings of the 24th International Conference on Data
  Engineering, {ICDE} 2008, April 7-12, 2008, Canc{\'{u}}n, M{\'{e}}xico},
  2008.

\bibitem{koudas@2006}
Nick Koudas, Chen Li, Anthony K.~H. Tung, and Rares Vernica.
\newblock Relaxing join and selection queries.
\newblock In {\em Proceedings of the 32nd International Conference on Very
  Large Data Bases, Seoul, Korea, September 12-15, 2006}, 2006.

\bibitem{quark-x}
Jyoti Leeka, Srikanta Bedathur, Debajyoti Bera, and Medha Atre.
\newblock \emph{Quark-X}: An efficient top-k processing framework for {RDF}
  quad stores.
\newblock In {\em Proceedings of the 25th {ACM} International on Conference on
  Information and Knowledge Management, {CIKM} 2016, Indianapolis, IN, USA,
  October 24-28, 2016}, 2016.

\bibitem{li@www2010}
Chengkai Li, Ning Yan, Senjuti~Basu Roy, Lekhendro Lisham, and Gautam Das.
\newblock Facetedpedia: dynamic generation of query-dependent faceted
  interfaces for wikipedia.
\newblock In {\em Proceedings of the 19th International Conference on World
  Wide Web, {WWW} 2010, Raleigh, North Carolina, USA, April 26-30, 2010}, 2010.

\bibitem{magliacane@iswc2012}
Sara Magliacane, Alessandro Bozzon, and Emanuele~Della Valle.
\newblock Efficient execution of top-k {SPARQL} queries.
\newblock In {\em The Semantic Web - {ISWC} 2012 - 11th International Semantic
  Web Conference, Boston, MA, USA, November 11-15, 2012, Proceedings, Part
  {I}}, 2012.

\bibitem{prob-empty-answer}
Davide Mottin, Alice Marascu, Senjuti~Basu Roy, Gautam Das, Themis Palpanas,
  and Yannis Velegrakis.
\newblock A probabilistic optimization framework for the empty-answer problem.
\newblock {\em {PVLDB}}, 6(14), 2013.

\bibitem{poulovassilis@iswc2010}
Alexandra Poulovassilis and Peter~T. Wood.
\newblock Combining approximation and relaxation in semantic web path queries.
\newblock In {\em The Semantic Web - {ISWC} 2010 - 9th International Semantic
  Web Conference, {ISWC} 2010, Shanghai, China, November 7-11, 2010, Revised
  Selected Papers, Part {I}}, 2010.

\bibitem{roy@cikm2008}
Senjuti~Basu Roy, Haidong Wang, Gautam Das, Ullas Nambiar, and Mukesh~K.
  Mohania.
\newblock Minimum-effort driven dynamic faceted search in structured databases.
\newblock In {\em Proceedings of the 17th {ACM} Conference on Information and
  Knowledge Management, {CIKM} 2008, Napa Valley, California, USA, October
  26-30, 2008}, 2008.

\bibitem{pbrj}
Karl Schnaitter and Neoklis Polyzotis.
\newblock Optimal algorithms for evaluating rank joins in database systems.
\newblock {\em {ACM} Trans. Database Syst.}, 35(1), 2010.

\bibitem{yago}
Fabian~M. Suchanek, Gjergji Kasneci, and Gerhard Weikum.
\newblock Yago: a core of semantic knowledge.
\newblock In {\em Proceedings of the 16th International Conference on World
  Wide Web, {WWW} 2007, Banff, Alberta, Canada, May 8-12, 2007}, 2007.

\bibitem{incmerge2005}
Martin Theobald, Ralf Schenkel, and Gerhard Weikum.
\newblock Efficient and self-tuning incremental query expansion for top-k query
  processing.
\newblock In {\em {SIGIR} 2005: Proceedings of the 28th Annual International
  {ACM} {SIGIR} Conference on Research and Development in Information
  Retrieval, Salvador, Brazil, August 15-19, 2005}, 2005.

\bibitem{top-x}
Martin Theobald, Ralf Schenkel, and Gerhard Weikum.
\newblock An efficient and versatile query engine for topx search.
\newblock In {\em Proceedings of the 31st International Conference on Very
  Large Data Bases, Trondheim, Norway, August 30 - September 2, 2005}, 2005.

\bibitem{theobald@vldb2004}
Martin Theobald, Gerhard Weikum, and Ralf Schenkel.
\newblock Top-k query evaluation with probabilistic guarantees.
\newblock In {\em (e)Proceedings of the Thirtieth International Conference on
  Very Large Data Bases, Toronto, Canada, August 31 - September 3 2004}, 2004.

\bibitem{tran@sigmod2010}
Quoc~Trung Tran and Chee{-}Yong Chan.
\newblock How to conquer why-not questions.
\newblock In {\em Proceedings of the {ACM} {SIGMOD} International Conference on
  Management of Data, {SIGMOD} 2010, Indianapolis, Indiana, USA, June 6-10,
  2010}, 2010.

\bibitem{DebEAQ}
Elena Vasilyeva, Thomas Heinze, Maik Thiele, and Wolfgang Lehner.
\newblock Debeaq - debugging empty-answer queries on large data graphs.
\newblock In {\em 32nd {IEEE} International Conference on Data Engineering,
  {ICDE} 2016, Helsinki, Finland, May 16-20, 2016}, 2016.

\bibitem{vasilyeva@adbis2014}
Elena Vasilyeva, Maik Thiele, Christof Bornh{\"{o}}vd, and Wolfgang Lehner.
\newblock Top-k differential queries in graph databases.
\newblock In {\em Advances in Databases and Information Systems - 18th East
  European Conference, {ADBIS} 2014, Ohrid, Macedonia, September 7-10, 2014.
  Proceedings}, 2014.

\bibitem{wagner@eswc2012}
Andreas Wagner, Duc~Thanh Tran, G{\"{u}}nter Ladwig, Andreas Harth, and Rudi
  Studer.
\newblock Top-k linked data query processing.
\newblock In {\em The Semantic Web: Research and Applications - 9th Extended
  Semantic Web Conference, {ESWC} 2012, Heraklion, Crete, Greece, May 27-31,
  2012. Proceedings}, 2012.

\bibitem{wang@IS2015}
Dong Wang, Lei Zou, and Dongyan Zhao.
\newblock Top-k queries on {RDF} graphs.
\newblock {\em Inf. Sci.}, 316, 2015.

\bibitem{xkg2016}
Mohamed Yahya, Denilson Barbosa, Klaus Berberich, Qiuyue Wang, and Gerhard
  Weikum.
\newblock Relationship queries on extended knowledge graphs.
\newblock In {\em Proceedings of the Ninth {ACM} International Conference on
  Web Search and Data Mining, San Francisco, CA, USA, February 22-25, 2016},
  2016.

\bibitem{yang@icde2016}
Shengqi Yang, Fangqiu Han, Yinghui Wu, and Xifeng Yan.
\newblock Fast top-k search in knowledge graphs.
\newblock In {\em 32nd {IEEE} International Conference on Data Engineering,
  {ICDE} 2016, Helsinki, Finland, May 16-20, 2016}, 2016.

\bibitem{yu@dexa2005}
Hailing Yu, Hua{-}Gang Li, Ping Wu, Divyakant Agrawal, and Amr {El Abbadi}.
\newblock Efficient processing of distributed top-\emph{k} queries.
\newblock In {\em Database and Expert Systems Applications, 16th International
  Conference, {DEXA} 2005, Copenhagen, Denmark, August 22-26, 2005,
  Proceedings}, 2005.

\end{thebibliography}
\end{document}